\documentclass[twocolumn]{pasj00}
\Received{2012/05/15}
\Accepted{2015/09/10}
\Published{$\langle$publication date$\rangle$}
\SetRunningHead{Y. Sugawara et al.}{{\it Suzaku} monitoring of WR~140
around periastron passage: An approach for quantifying the wind parameters}
\usepackage{color}
\usepackage{times}

\begin{document}

\title{{\it Suzaku} monitoring of the Wolf-Rayet binary WR~140 around
periastron passage: An approach for quantifying the wind parameters}

\author{Yasuharu~\textsc{Sugawara},\altaffilmark{1}
Yoshitomo~\textsc{Maeda},\altaffilmark{2,3}
Yohko~\textsc{Tsuboi},\altaffilmark{1}
Kenji~\textsc{Hamaguchi},\altaffilmark{4,5}
Michael~\textsc{Corcoran},\altaffilmark{4,5}
A.~M.~T.~\textsc{Pollock},\altaffilmark{7,8}
Anthony~F.~J.~\textsc{Moffat},\altaffilmark{9}
Peredur~M.~\textsc{Williams},\altaffilmark{10} 
Sean~\textsc{Dougherty},\altaffilmark{11} and
Julian~\textsc{Pittard},\altaffilmark{12}}

\altaffiltext{1}{Department of Physics, Faculty of Science \&
Engineering, Chuo University, 1-13-27 Kasuga, Bunkyo, Tokyo 112-8551}
\altaffiltext{2}{Institute of Space and Astronautical Science, Japan
Aerospace Exploration Agency 3-1-1 Yoshinodai, Sagamihara, Kanagawa
229-8510}
\altaffiltext{3}{Department of Space and Astronautical Science, SOKENDAI
(The Graduate University for Advanced Studies), 3-1-1 Yoshinodai,
Chuo-ku, Sagamihara, Kanagawa 252-5210}
\altaffiltext{4}{CRESST and X-ray Astrophysics Laboratory NASA/GSFC,
Greenbelt, MD 20771, USA}
\altaffiltext{5}{Department of Physics, University of Maryland,
Baltimore County, 1000 Hilltop Circle, Baltimore, MD 21250, USA}
\altaffiltext{6}{Universities Space Research Association, 7178 Columbia
Gateway Drive, Columbia, MD 21046, USA}
\altaffiltext{7}{European Space Agency, XMM-Newton Science Operations
Centre, European Space Astronomy Centre, Apartado 78, Villanueva de la
Ca$\tilde{n}$ada, 28691 Madrid, Spain}
\altaffiltext{8}{Department of Physics and Astronomy, University of
Sheffield, Sheffield S3 7RH, England}
\altaffiltext{9}{D$\acute{e}$partement de Physique, Universit$\acute{e}$
de Montr$\acute{e}$al, Succursale Centre-Ville, Montr$\acute{e}$al, QC
H3C 3J7, and Centre de Recherche en Astrophysique du Qu\'ebec, Canada}
\altaffiltext{10}{Institute for Astronomy, Royal Observatory, Blackford
Hill, Edinburgh EH9 3HJ, Scotland}
\altaffiltext{11}{National Research Council of Canada, DRAO, Penticton}
\altaffiltext{12}{School of Physics and Astronomy, The University of
Leeds, Leeds LS2 9JT}

\email{sugawara@phys.chuo-u.ac.jp}

\maketitle

\begin{abstract} 
 {\it Suzaku} observations of the Wolf-Rayet binary WR 140
 (WC7pd$+$O5.5fc) were made at four different times around periastron
 passage in 2009 January. The spectra changed in shape and flux with the
 phase. As periastron approached, the column density of the low-energy
 absorption increased, which indicates that the emission from the
 wind-wind collision plasma was absorbed by the dense W-R wind. The
 spectra can be mostly fitted with two different components: a warm
 component with $k_{\rm B}T=$0.3--0.6 {\rm keV} and a dominant hot
 component with $k_{\rm B}T \sim$3 {\rm keV}.  The emission measure of
 the dominant, hot component is not inversely proportional to the
 distance between the two stars. This can be explained by the O star
 wind colliding before it has reached its terminal velocity, leading to
 a reduction in its wind momentum flux. At phases closer to periastron,
 we discovered a cool plasma component in a recombining phase, which is
 less absorbed.  This component may be a relic of the wind-wind
 collision plasma, which was cooled down by radiation, and may represent
 a transitional stage in dust formation.

\end{abstract}

\section{Introduction}
Mass-loss is one of the most important and uncertain parameters in the
evolution of a massive star. There are several methods for determining
mass-loss rates (e.g., via radio continuum flux, radiative transfer and
polarization variation in close binaries).  It has become increasingly
recognized (e.g., \cite{puls06}) that smooth-wind models, based on
density-squared diagnostics, overestimate clumped wind mass-loss rates
by up to an order of magnitude.

Another important parameter for massive stars is the wind acceleration.
For most radiatively-driven stellar-wind models, a velocity law
$v(r)=v_{\infty}(1-R/r)^{\beta}$ with $\beta=1$ is assumed.  Here,
$v_{\infty}$ and $\it{R}$ are the terminal wind-velocity and stellar
radius, respectively.  In this model, usually the initial wind velocity
$v_0$ is neglected, since it is thought to be $\sim$1\% of $v_{\infty}$
and the wind-collision X-rays are formed relatively far out in the
wind. On the other hand, some optical observations have revealed a high
value of $\beta$ for Wolf-Rayet (W-R) stars (e.g., 20~${\rm R}_{\odot} <\beta
R<80~{\rm R}_{\odot}$; \cite{lepine99}, which lead to values
$\beta~\gg$~1 for normal W-R radii).

\begin{table*}[ht!]
 \begin{center}
  \caption{{\it Suzaku} Observation Log.}\label{tb:t2-1}
  \begin{tabular}{lccllllrr}
   \hline
   Obs.  & Observation  &Observation
   &$t_{\rm{exp}}$\footnotemark[$*$] &$\dot{C}_{\rm
   XIS/FI}$\footnotemark[$\dagger$]& $\dot{C}_{\rm
   XIS/BI}$\footnotemark[$\dagger$]&$\dot{C}_{\rm
   HXD/PIN}$\footnotemark[$\dagger$]&Orbital &{\it D}\footnotemark[$\ddagger$] \\
           & Start [UT]        &End [UT]          &[ks]
		   &[cps]&[cps]&[cps]&Phase\footnotemark[$\ddagger$]& [AU]\\
   \hline
   A  & 2008-04-09 05:33:13 & 2008-04-09 17:20:18 & 21.6 &
		       2.186$\pm$0.007&2.68$\pm$0.01&0.024$\pm$0.005&
                       2.904 & 13.79\\
   B  & 2008-12-12 10:27:51 & 2008-12-13 13:15:20 & 52.8 &
                       2.262$\pm$0.005&2.415$\pm$0.007&0.029$\pm$0.003&
			       2.989 & 3.12\\
   C  & 2009-01-04 08:36:00 & 2009-01-05 10:12:18 & 47.3 &
		       0.661$\pm$0.003&0.644$\pm$0.004&0.008$\pm$0.003&
                       2.997 & 1.73\\
   D  & 2009-01-13 12:59:45 & 2009-01-15 12:00:13 & 89.4 &
		       0.193$\pm$0.001&0.178$\pm$0.002&0.005$\pm$0.002&
                       3.000 & 1.53\\
   \hline
   \multicolumn{9}{@{}l@{}}{\hbox to 0pt{\parbox{170mm}{\footnotesize
   \par\noindent
   \footnotemark[$*$] Net exposure of the XIS.\\
   \footnotemark[$\dagger$] $\dot{C}$ shows net count rate in counts per
   second (cps). The bandpasses are 0.4--10 keV for the FI and BI
   detectors and 15--50 keV for the PIN detectors.\\
   \footnotemark[$\ddagger$] According to \citet{monnier11}. {\it D}
   shows binary separation.
  }\hss}}
  \end{tabular}
 \end{center}
\end{table*}

\begin{figure}[ht!]
 \begin{center}
  \FigureFile(85mm,85mm){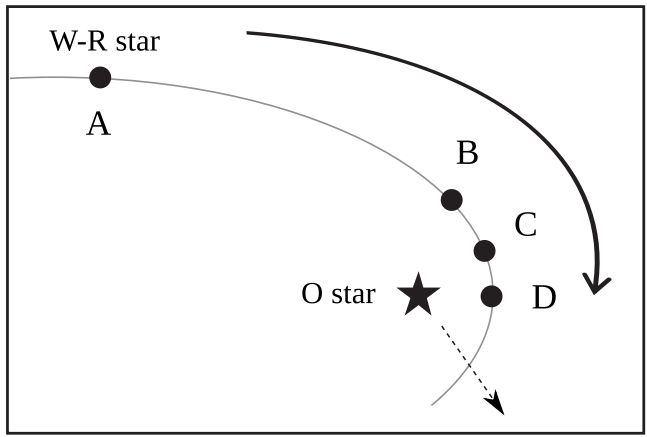}
 \end{center}
 \caption{Schematic view of the orbit of the WR~140
 system. The filled circles show relative positions of the W-R star
 during observations A--D. The dashed arrow shows the line of sight to Earth.}
\label{fg:f2-1} 
\end{figure}

While the mass-loss rate and the acceleration parameter $\beta$ have
been measured using the radio/IR continuum flux or line spectral
analysis at optical/IR wavelengths, X-ray wavelength could be an
independent window to approach these parameters.  Colliding wind
binary is a good target, having variable X-ray spectra with orbital
phase.  The X-ray is emitted from the wind-shocked region, which is
strongly dependent on the ram-pressure balance between the two
hypersonic winds.  The shocked plasmas, which have temperatures of
10$^{7}$--10$^{8}$~{\rm K}, are frequently observed, and the high
absorption column of 10$^{22}$--10$^{23}$ H cm$^{-2}$ are reported
(cf. \cite{schild04}).  The temperature should reflect on the wind
velocity, and the absorption column indicates the dense W-R wind
\citep{koyama94}, i.e. mass-loss rate of the W-R star
(cf. \cite{pollock05}). The X-ray luminosity is highly dependent on the
separation between the stars of the binary, the mass-loss rates, and
wind velocities (\cite{stevens92}; \cite{usov92}).  If we know the
orbital parameters of the binary, the X-ray luminosity at each orbital
phase should depend on the mass-loss rates and wind-acceleration
parameters.

WR~140 (HD 193793) is considered as the textbook example of a colliding
wind binary. The star has been classified as a WC7pd$+$O5.5fc binary
system whose stellar masses are ${\it M}_{\rm {WR}}$ = 16 $M_{\odot}$
and ${\it M}_{\rm {O}}$ = 41 $M_{\odot}$ by the optical spectroscopic
monitoring \citep{fahed11}. Its orbit and distance have been well
determined with ${\it P}_{\rm {orb}}$ = 2896.35 days, {\it i} =
119.6$\degree$, {\it e} = 0.8964 and {\it d} = 1.67 kpc by detailed IR
astrometric study \citep{monnier11}. Radio non-thermal (synchrotron)
emission from WR~140 was reported that may arise from a wind-wind
collision zone (e.g., \cite{williams90}; \cite{white95b};
\cite{dougherty05}). \citet{pittard06} proposed a radio, X-ray and
$\gamma$-ray non-thermal emission model.  In their model, there are some
cases where inverse Compton emission dominates the high energy X-ray
emission. \citet{pittard06} predicted that {\it Suzaku} would detect
such non-thermal X-ray emission.

WR~140 is a bright X-ray source (e.g., \cite{pollock87};
\cite{williams90}; \cite{zhekov00}; \cite{pollock02}; \cite{pollock05}).
During the previous periastron passage, {\it Rossi-XTE} showed a drop in
X-ray flux before periastron \citep{pollock05}.  We pursued this via
{\it Suzaku} monitoring during the following periastron passage in 2009.
These observations cover the energy range 0.1--20 keV.

\section{Observations and Data Reduction}
The {\it Suzaku} X-ray observatory \citep{mitsuda07} is equipped with
two kinds of instruments; the XRT (X-Ray Telescope,
\cite{serlemitsos07}) $+$ XIS (X-ray Imaging Spectrometer:
\cite{koyama07}) system, sensitive to X-rays between 0.3--12~{\rm keV},
and the HXD (Hard X-ray Detector: \cite{kokubun07,takahashi07})
sensitive to X-rays above 10~{\rm keV}.  {\it Suzaku} observed WR~140
four times around periastron passage in 2009 January. Sequence numbers
of the data are 403030010, 403031010, 403032010 and 403033010. Logs of
these four observations, labeled as A, B, C and D, are summarized in
table~\ref{tb:t2-1} and figure ~\ref{fg:f2-1}.  The total exposure of
these observation was $\sim$210 {\rm ks}.

The XIS is composed of four X-ray CCD (XIS0--3) arrays with
1024$\times$1024~pixel formats, each of which is mounted at the focal
plane of an individual XRT.  The XRT+XIS system covers a field of view
of $\sim$18\arcmin$\times$18\arcmin.  The XIS1 has a back-side
illuminated (BI) CCD chip while the remaining active sensors (XIS0 and
XIS3) have front-side illuminated (FI) chips.  The BI and FI chips are
superior to each other in the soft and hard band responses respectively.
During the observations, the XISs were operated in the normal clocking
mode with the default frame time (8~{\rm s}). WR~140 was placed at the
HXD nominal position.

The HXD is a non-imaging instrument that consists of a 4$\times$4 array
of detectors (well units) and 20 surrounding crystal scintillators for
active shielding.  Each well unit consists of four Si PIN diodes (PIN),
sensitive to X-rays between 10--70~{\rm keV}, and four GSO/BGO phoswitch
counters (GSO) for detecting photons with energies in the range
40--600~{\rm keV}. We used only PIN data to constrain the emission above
10~keV due to the high background of the GSO instrument.

\begin{figure*}[ht!]
 \begin{center}
  \FigureFile(170mm,85mm){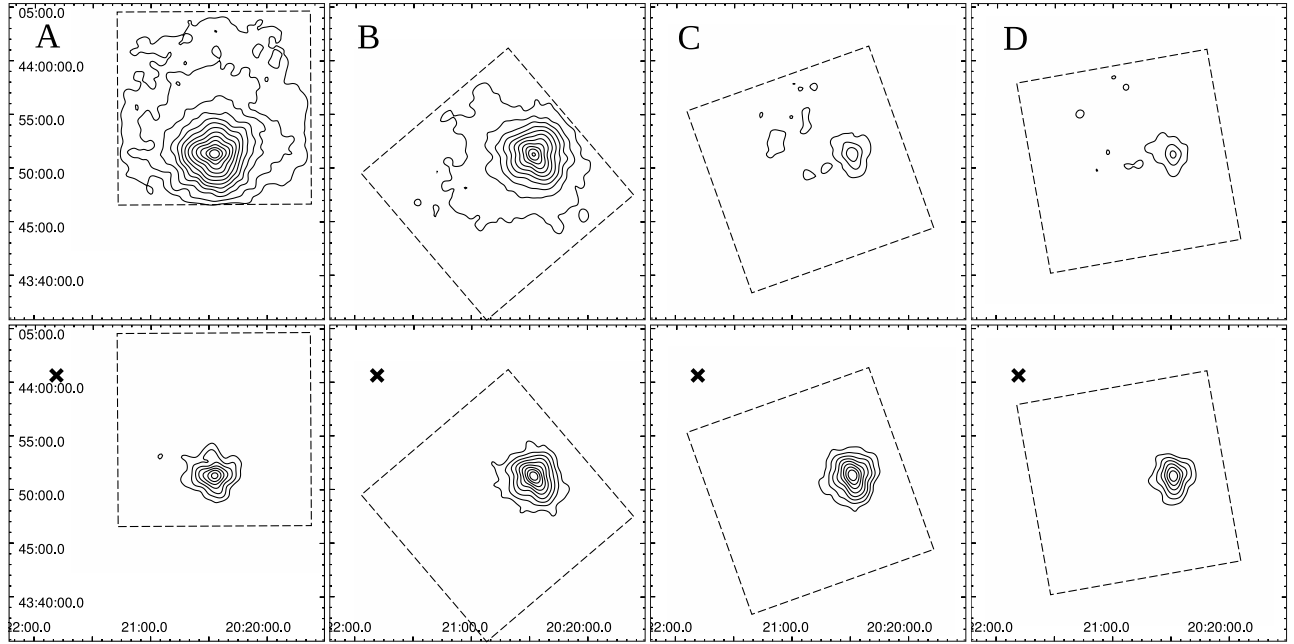}
 \end{center}
 \caption{{\it Suzaku} XIS (0+1+3) contour images of WR~140. Top panels:
0.4--1.6 {\rm keV} band. Bottom panels: 7--10 {\rm keV} band.  The
contour levels are for soft band, 0.04, 0.08, 0.16, 0.32, 0.64, 1.28,
2.56, 5.12, 10.24, 20.48, 40.96, 81.92 and for hard band, 0.025, 0.04,
0.064, 0.102,0.164, 0.262, 0.419, 0.671, 1.074 in units of counts
ks$^{-1}$ pixel$^{-1}$. In the soft band images of the observation C and
D, some weak sources are located in the west region of WR~140. Since the
flux of these sources varied between observations C and D, they are
considered as the variable point sources. The dashed lines show the XIS
field of view. The crosses show the location of IGR J20216+4359.}
\label{fg:f1}
\end{figure*}

\begin{figure*}[ht!]
 \begin{center}
  \FigureFile(170mm,170mm){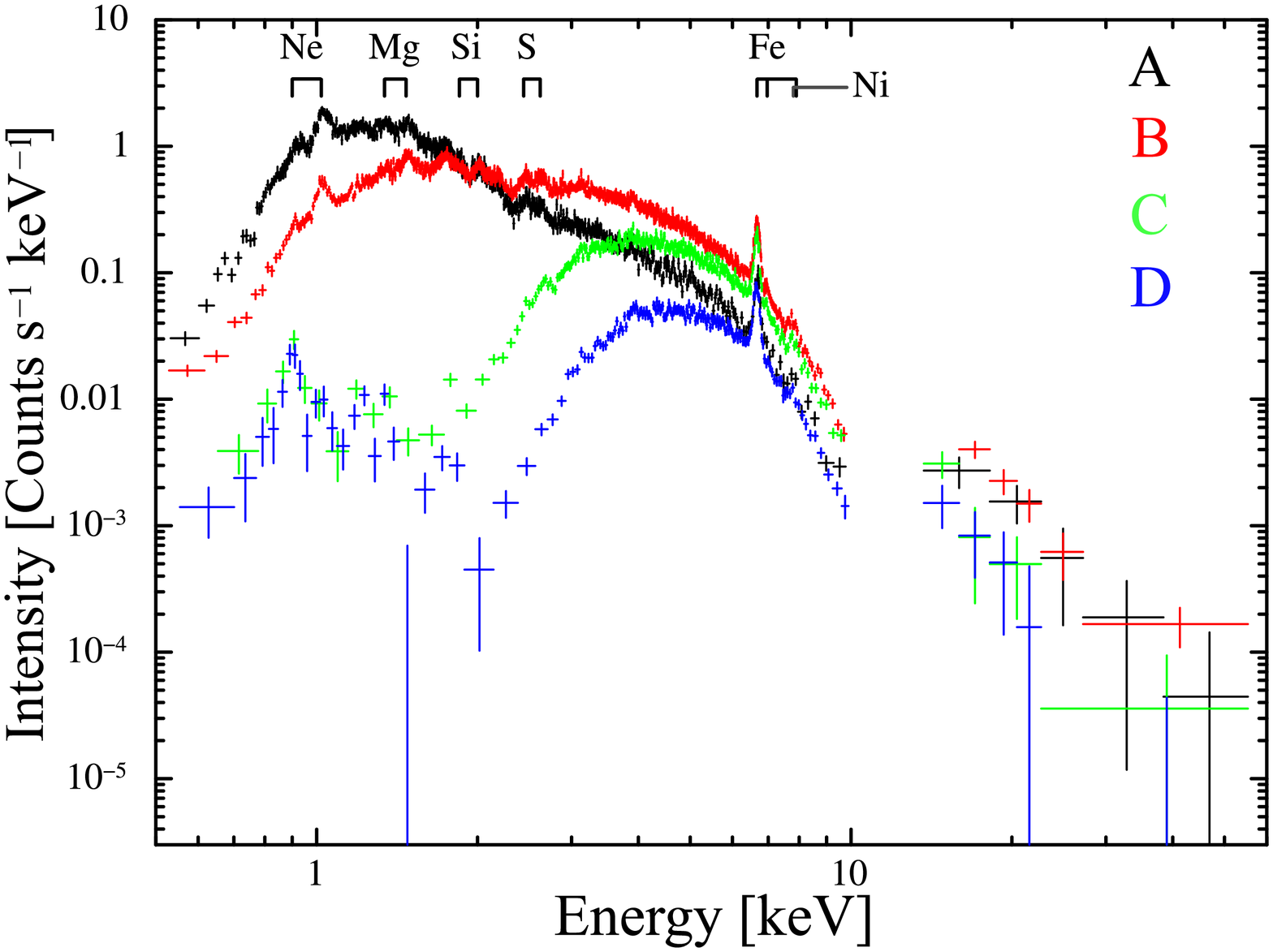}
 \end{center}
 \caption{Background subtracted XIS/FI and HXD/PIN spectra of
 WR~140.} 
\label{fg:f3-2}
\end{figure*}

We reduced the data using
HEAsoft\footnote{http://heasarc.gsfc.nasa.gov/docs/software/lheasoft/}
version 6.16.0 and calibration versions (CALDBVER) xis20090402,
xrt20080709 and hxd20090402.  We screened out XIS and HXD events
obtained during (1) the South Atlantic Anomaly (SAA) passage, (2) the
low geomagnetic cut-off rigidity (6~{\rm GV} for the XIS and 8~{\rm GV}
for the HXD), (3) low elevation angles from the Earth rim
($<$10$^{\circ}$ for the XIS and $<$5$^{\circ}$ for the HXD) and the
sun-lit Earth rim ($<$20$^{\circ}$), and (4) telemetry saturation.  We
used XSPEC\footnote{http://heasarc.gsfc.nasa.gov/xanadu/xspec/} version
12.5.1 for spectral analysis.

\section{Data Analysis and Results}

\begin{figure*}[ht!]
 \begin{center}
  \FigureFile(160mm,70mm){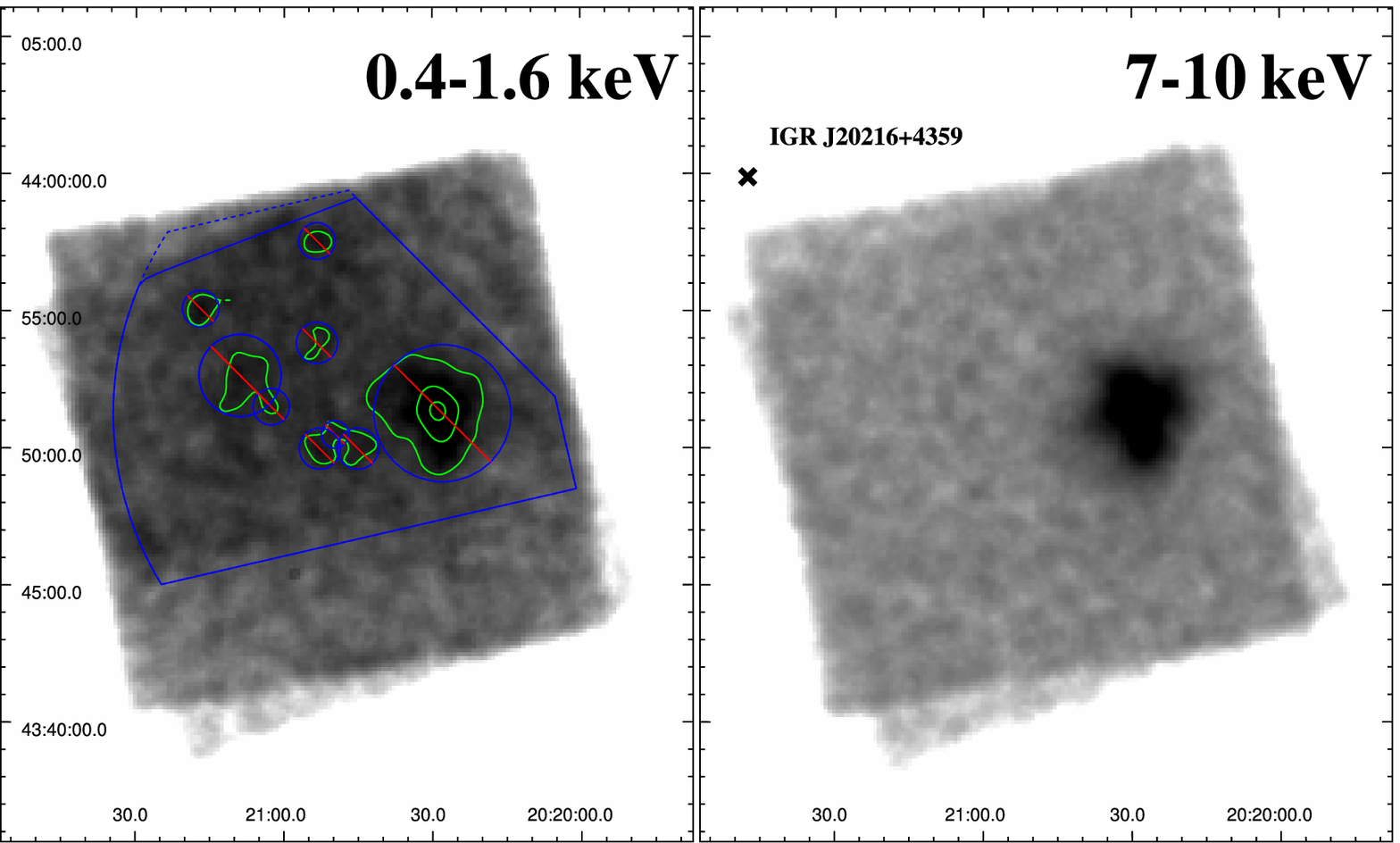}
 \end{center}
 \caption{{\it Suzaku} XIS (0+1+3) images with a log color scale
 obtained at observations C and D. We averaged the images of
 observations C and D. Left panel shows 0.4--1.6~{\rm keV} band, and
 right panel shows 7--10~{\rm keV} band. Images have been smoothed with
 a Gaussian function with $\sigma=$2.5 image pixels ($\sim$~21
 arcsec). The contour levels are 0.039, 0.073, 0.138 in units of counts
 ks$^{-1}$ pixel$^{-1}$. We used the arc region as the background of the
 cool component excluding nine circle areas (see the text for the
 detail). The solid and dashed outer lines show the background regions
 for observation C and D, respectively. The XIS has a spatial resolution
 of about 2~arcmin in a half-power diameter.}  \label{fg:3-3}
\end{figure*}
\begin{figure*}[!htb]
 \begin{center}
  \FigureFile(140mm,60mm){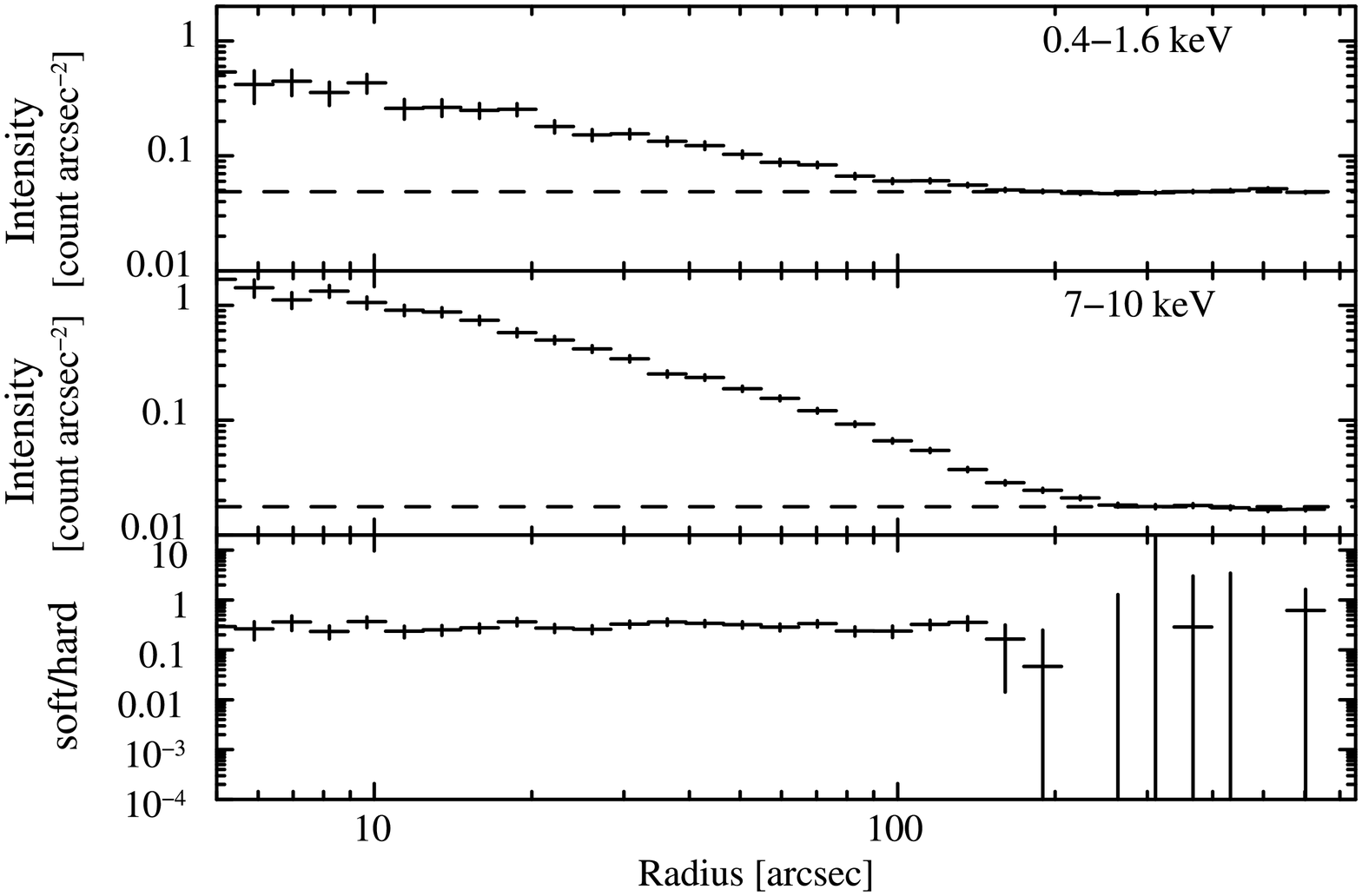}
 \end{center}
 \caption{Radial profiles which are centered on WR~140. The data are
 summed up for XIS-0, 1 and 3. Only observation D is used. Top:
 0.4--1.6 {\rm keV} (soft). Center: 7--10~{\rm keV} (hard).  Dashed
 lines show the count rates of our assumed background level that
 consists of the Cygnus superbubble and the NXB. Bottom: The ratio of
 the soft to the hard energy-band data after the constant background is
 subtracted.} \label{fg:3-3-2}
\end{figure*}

\subsection{Spectral Characteristics}
\label{subsec:spectral_chara}

Figure~\ref{fg:f1} shows the XIS images in the soft (0.4--1.6~{\rm keV}) band
and the hard (7--10~{\rm keV}) band. In all the observations, the XISs
detected an X-ray source at the position of WR~140. In order to capture the
spectral characteristics of WR~140, we extracted the source events from a
circle with 4~arcmin radius centered on the X-ray peak, which gives a
maximum signal-to-noise ratio in the 0.4--10~{\rm keV} band. We
extracted background events from the entire XIS field of view, excluding
the region within 5 arcmin from the X-ray peak and calibration source
regions.
WR~140 did not show any significant X-ray variation during each
observation, so that we generated time-averaged spectra for each
observation. We generated XIS detector responses (RMF) and XRT
effective area tables (ARF) with {\tt xisrmfgen} and {\tt xissimarfgen}
\citep{ishisaki07} and averaged two FI spectra and responses with the
HEAsoft {\tt mathpha}, {\tt addrmf} and {\tt addarf}. The FI spectra of
observations A and B, C, D were grouped to a minimum of 100 and 200
counts per bin with {\tt FTOOL grppha}, respectively. Similarly, the BI
spectra of observations A, B, C and D were grouped to a minimum of
100, 200, 150 and 200 counts per bin, respectively. In spectral fits, we
ignored XIS bins around the Si K-edge between 1.8--2.0~{\rm keV} because
of the calibration uncertainty.

We noticed that the O-Ly$\alpha$, Ne-Ly$\alpha$ and Mg-Ly$\alpha$ lines
in the BI spectra of observations A and B deviate significantly from
those lines in the Astrophysical Plasma Emission Database
(APED)\footnote{See http://www.atomdb.org/ for details.}.  The
FI spectra did not show such deviation, and are consistent within
$\sim$5~{\rm eV} with the values in the database. This is suggesting
that the energy gain of the BI chip is not calibrated sufficiently.
Center energies of these lines, measured with fits by Gaussian
functions, deviate by $-$5$\pm$3~{\rm eV} and $-$20$\pm$3~{\rm eV} for
observations A and B, respectively.  We therefore offset those
spectra by the best-fit values. We did not make gain offsets for
the spectra of observations C and D, since they did not show those emission
lines.

We subtracted from the pipeline-processed HXD/PIN spectra non-X-ray
background (NXB) simulated with the tuned NXB model (LCFITDT,
\cite{fukazawa09}) and the typical Cosmic X-ray Background (CXB)
measured with the {\it HEAO-1} \citep{boldt87} \footnote{See
http://heasarc.gsfc.nasa.gov/docs/suzaku/analysis/pin\_cxb.html}.  We
generated the PIN detector response with {\tt hxdarfgen}. The PIN
spectra also were binned with {\tt grppha} \footnote{Observation A: group
0 35 6 36 71 12 72 101 30 102 236 45 237 255 19; Observation B: group 0
59 6 60 71 12 72 146 75 147 236 45 237 255 19; Observation C: group 0 47
6 48 59 12 60 146 87 147 236 90 237 255 19; Observation D: group 0 59 6
60 146 87 147 236 45 237 255 19}.

The XIS spectra of WR~140 (figure~\ref{fg:f3-2}) show emission lines
from highly ionized ions of neon, magnesium, silicon, sulfur and iron.
Earlier observations of WR~140 ({\it Ginga}: \cite{koyama90}, {\it
ASCA}: \cite{koyama94,zhekov00}, {\it Chandra}: \cite{pollock05}, {\it
XMM-Newton}: \cite{debecker11}) also found these lines in the spectra,
suggesting optically thin plasma thermalized by wind-wind collision.
The low energy cut-off --- the signature of absorption by neutral matter
or weakly ionized plasma --- drastically increased from observation A to
D.  High density and/or large atomic number are needed for large
absorption. In WR~140, the mass-loss rate of the O-star is an order of
magnitude smaller than that of the W-R star. In addition, the W-R wind
have higher metal abundances than the O-star wind. Therefore, it is well
known that the absorption by the W-R star is dominant (e.g.,
\cite{koyama94}). The spectral variation like this was reported by
\citet{zhekov00}. We call this dominant emission ``variable component''
(see $\S$~\ref{subsec:variable_comp}). During observations C and D when
soft X-rays from the wind-wind collision are suppressed by absorption
from the wind of the W-R star, we detected another emission component
below 2~keV.  This component did not change apparently between these
observations.  We call this soft emission ``cool component'' (see
$\S$~\ref{subsec:cool_comp}).

The HXD/PIN detected significant X-ray emission in the 15--50~{\rm keV}
range. The flux seems to change from observation B to D (see
figure~\ref{fg:f3-2}), which seems to correlate with the variation below
10~{\rm keV}. However, we should pay attention to other hard X-ray
source IGR J20216+4359, which is located at 17.35 arcmin away from
WR~140 \citep{bikmaev08}.  The location is shown in figure~\ref{fg:f1}
\& \ref{fg:3-3}. Because of a potential for the contamination from IGR
J20216+4359 in the field of view of PIN/HXD, XIS and PIN spectra are
analyzed separately.

\subsection{Cool Component}
\label{subsec:cool_comp} 

We found the cool component in the spectra of observations C and D (see
$\S$~\ref{subsec:spectral_chara}).  In this subsection, we report the
analysis results of the cool component. First, we check the spatial
structure of the cool component. Then, we compare the spectral features
of diffuse emission around WR~140 and the cool component. Finally, we
fit the spectra of the cool component.

\subsubsection{Radial extent of the cool component}
\label{subsubsec:cool_dist}

WR~140 is located in the Cygnus region that is pervaded by diffuse X-ray
emission from a large ring-like structure, whose diameter is
$\approx13^\circ$. This diffuse structure is known as the Cygnus
superbubble (CSB). We then figure out the spatial structure of the cool
component before the spectral analysis.

Figure~\ref{fg:3-3} shows {\it Suzaku} images in the 0.4--1.6~{\rm keV}
and 7--10~{\rm keV} band, combined from observations C and D. In the
soft band, the diffuse emission with some degree of inhomogeneity is
seen.  The north and east areas were brighter than the south area by a
factor of two in count rate. Some transient sources were identified in
observations C and D.

\begin{figure}[!ht]
 \begin{center}
  \FigureFile(70mm,70mm){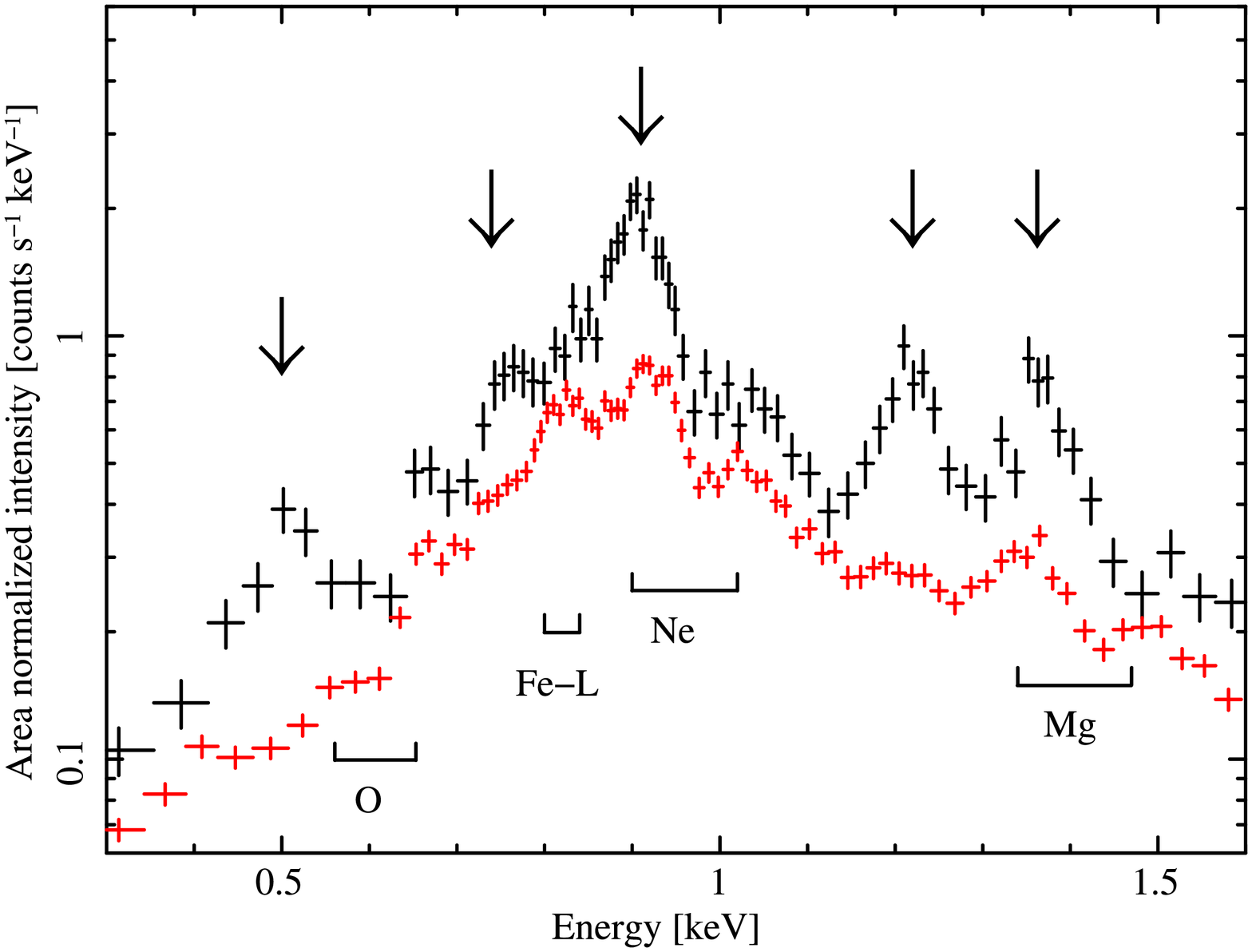}
 \end{center}
 \caption{Comparison of the spectra of the source (black) and background
 (red) regions. The data of XIS-0, 1 and 3 for observations C and D were
 summed to make the spectra.  Downward-pointing arrows indicate
 prominent excess of the source spectrum over the background.}
 \label{fg:3-3-3}
\end{figure}

Figure~\ref{fg:3-3-2} shows the radial profiles of the {\it Suzaku} XIS
images for observation D. The images are taken with XIS-0, 1, and 3, and
summed up. In figure~\ref{fg:3-3-2}, the dashed line in each panel shows
the background level determined by the fitting of the data between
200--600 arcsec form WR~140 with a constant model.
The bottom panel in figure~\ref{fg:3-3-2} shows the ratio of the soft
band count rate and the hard band count rate, after background
subtraction. The ratio is nearly flat along the radial axis. The point
spread functions of the XISs do not significantly show the energy
dependency\citep{serlemitsos07}. Since the emission from the point
source WR~140 dominated the XIS count rate in the hard band, the
constant ratio indicates that the cool component is also point-like and
associated with WR~140.

\subsubsection{Spectral features}
\label{subsubsec:cool_spec} In this subsubsection, we compare the
spectral features of the background diffuse emission around WR~140 and
the cool component.  Since WR~140 lies in the north area of the diffuse
structure, we adopted the north and east regions of WR~140 as the
background region for observations C and D. We excluded the transient
sources from the background region, as shown in figure~\ref{fg:3-3}.  In
order to improve the signal-to-noise ratio of the spectra in the
0.4--1.6 {\rm keV} band, we extracted source events within 2~arcmin from
the X-ray peak. We co-added the XIS-0, 1 and 3 spectra
for observations C and D and ignored spectral bins above 1.6~{\rm keV}.

Figure~\ref{fg:3-3-3} shows the area-normalized spectra of the source
and the background region. NXB subtraction were not made for both of the
spectra.  The spectra of the background region was fairly complex, which
had O-Ly$\alpha$, Ne-Ly$\alpha$ and Mg-Ly$\alpha$ lines and the
structures of He-like O and Ne.  On the other hand, the source spectrum
shows large excess from the background. Especially, the structures,
which are shown with arrows in figure~\ref{fg:3-3-3}, are unique to the
source spectrum, and cannot be made by an inappropriate background
subtraction. At least the 1.2~{\rm keV} peaked structure in these unique
structures could not been reproduced by collisional equilibrium (CE)
plasma model. The detail is discussed in $\S$~\ref{subsubsec:cool_fit}.

\begin{table}[!ht]
\caption{Results of spectral fitting for cool component.}
\label{tb:t3-4} 
\renewcommand{\footnoterule}{}
\begin{center}
\begin{tabular}{l l c}
\hline\vspace{-3mm}\\
Components &Parameters&\\
\hline
Absorption& $N_{\mathrm{H}}^{\rm cool}$ (10$^{21}$ $\mathrm{cm}^{-2}$)&3.8$^{+2.0}_{-1.7}$\vspace{1mm}\\
vrnei& $k_{\rm{B}}T^{\rm cool}$\footnotemark[$*$] (keV)&0.021$^{+0.006}_{-0.005}$\vspace{1mm}\\
& $k_{\rm{B}}T^{\rm cool}_{\rm init}$\footnotemark[$\dagger$] (keV)&$>$0.58\vspace{1mm}\\
&N (solar)& $<$47\vspace{1mm}\\
&O (solar)&  9.1$^{+7.0}_{-4.6}$\vspace{1mm}\\
&Ne (solar)& 17.1$^{+24.1}_{-11.5}$\vspace{1mm}\\
&Mg (solar)& 3.4$^{+7.4}_{-3.2}$\vspace{1mm}\\
& $\tau_{\rm u}$\footnotemark[$\ddagger$](10$^{11}$ s
     cm$^{-3}$)&2.3$^{+0.7}_{-0.5}$\vspace{1mm}\\
& norm \footnotemark[$\S$] (10$^{-5}$)& 7.5$\pm0.7$\ \vspace{1mm}\\
\hline
&$F_{\mathrm{X}}$ \footnotemark[$\|$](10$^{-13}$ $\mathrm{erg}$
     $\mathrm{cm}^{-2}$ $\mathrm{s}^{-1}$) &  1.1\vspace{1mm}\\
 & $L_{\mathrm{X}}$ \footnotemark[$\#$](10$^{32}$ $\mathrm{erg}$ $\mathrm{s}^{-1}$)& 2.0\vspace{1mm}\\
& $\chi^{2}$/dof&98/83\\
\hline
\multicolumn{3}{@{}l@{}}{\hbox to 0pt{\parbox{190mm}{\footnotesize
   \par\noindent
Notes. 
We used the spectra of observations C and D, by averaging them. We\\
 did a simultaneous fitting of the FI and BI
spectra. We adopted the model \\{\tt
 TBabs} * {\tt vrnei}. The best-fit model is shown in figure 7 {\it right} panel with\\ a solid line. Errors
 quoted are the 90\% confidence interval.\\
 \footnotemark[$*$] Plasma temperature\\
 \footnotemark[$\dagger$] Initial plasma temperature\\
 \footnotemark[$\ddagger$] Ionization timescale.\\
 \footnotemark[$\S$] Normalization constant defined as ${\it E.M.} \times
 10^{-14}(4 \pi {\it d}^{2})^{-1}$, where ${\it E.M.}$ \\
is the emission measure in $\rm{cm}^{-3}$ and {\it d} is the distance in {\rm cm}.\\
 \footnotemark[$\|$]The observed flux (0.4--1.6 {\rm keV})\\
 \footnotemark[$\#$]The absorption-corrected luminosity (0.4--1.6
 {\rm keV}), which we calculated \\by assuming a distance of 1.67 {\rm kpc}.\\
    }\hss}}
\par\noindent
\end{tabular}
\end{center}
\end{table}

\begin{figure*}[!ht]
 \begin{center}
  \FigureFile(160mm,70mm){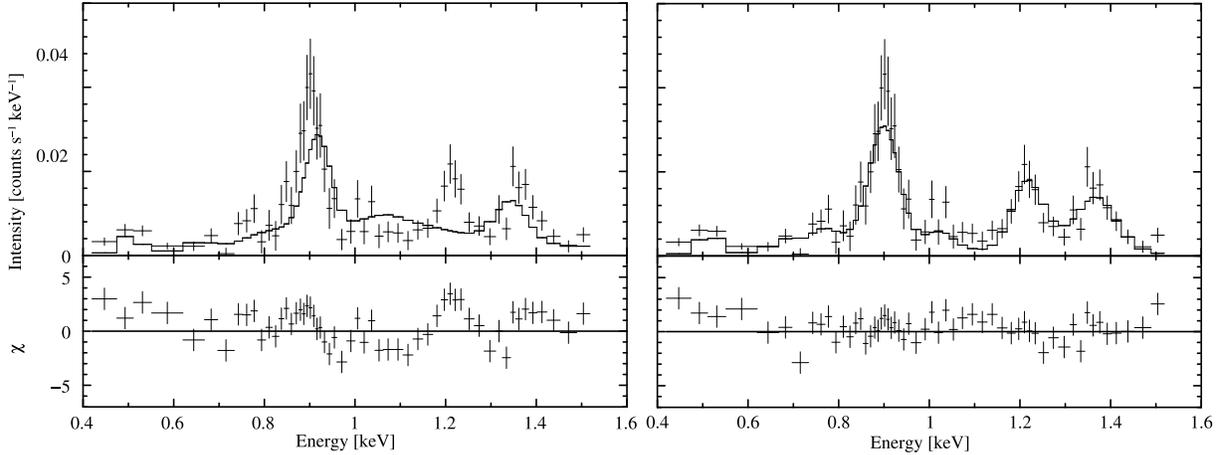}
 \end{center}
 \caption{The background subtracted FI spectra below 1.6 {\rm keV},
 where the spectra of observations C and D are averaged. The spectrum
 was grouped to a minimum of 40 counts per bin.  {\it Left}: Fit with
 the {\tt TBabs} * {\tt vAPEC} model. {\it Right}: Fit with the {\tt
 TBabs} * {\tt vrnei} model. The solid lines show each model. The lower
 panels show the residuals of the data from the model. See also
 table~\ref{tb:t3-4} for the best-fit parameters.}  \label{fg:3-4}
\end{figure*}

The observed surface brightness of background region was $4 \times
10^{-14}~\mathrm{erg}~%
\mathrm{s}^{-1}~\mathrm{cm}^{-2}~\mathrm{arcmin}^{-2}$ (0.5--2~{\rm keV}).
This value was obtained from a fit of the NXB subtracted BI spectrum for
observation D by a thin thermal plasma model ({\tt APEC}:
\cite{smith01}).  We adopted the ancillary response file for a circular
flat field sky with a 20 arcmin radius.  This observed surface
brightness is not inconsistent with the reported flux within an order of
magnitude \citep{zhekov00}.

\subsubsection{Spectral fitting}
\label{subsubsec:cool_fit} 

In this subsubsection, we fit the spectra of the cool component in the
0.4--1.6~{\rm keV} band with two models. 
First, we fitted the FI and BI background subtracted spectra using a
model of a one-temperature CE plasma ({\tt vAPEC}) emission suffering
absorption by cold matter with the interstellar medium (ISM) abundances
({\tt TBabs}: \cite{wilms00}).
As for the CE plasma, the elemental abundance were fixed at the best-fit
values for the variable component (see $\S$~\ref{subsubsec:globalfitAB}
and table~\ref{tb:t3-2}) except for N, Ne and Mg. The fitting result
(figure~\ref{fg:3-4} {\it left}) yields a plasma temperature of $k_{\rm
B}T^{\rm cool} \sim 0.2$ {\rm keV} and a column density $N_{\rm H}^{\rm
cool}\sim7\times$ 10$^{21}$ cm$^{-2}$ ($\chi^{2}$/dof $=$ 163/50).

\begin{figure}[ht!]
 \begin{center}
  \FigureFile(85mm,85mm){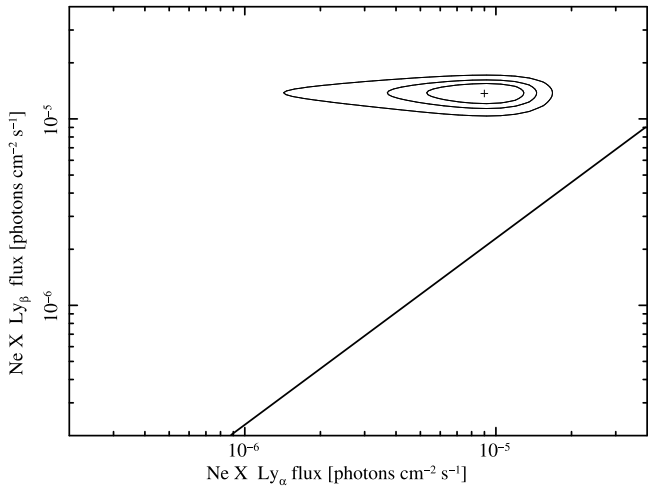}
 \end{center}
\caption{The confidence contour map of the normalizations of the two
Gaussian lines (Ne-Ly$\alpha$ and Ly$\beta$). The best-fit parameters
were obtained by a simultaneous fitting of FI and BI spectra that were
made by averaging the data of observations C and D. The contours are
at 68\%, 90\%, and 99\% confidence levels for two interesting
parameters.  The solid straight line indicates the line ratio of
Ly$\alpha$ to Ly$\beta$, based on the {\tt APEC} model.}
\label{fg:f6-3}
\end{figure}

This model still leaves line-shaped residuals at 1.21$\pm 0.01$ {\rm
keV}.  In the emission line database APED, the 1.21 {\rm keV} line
corresponds to Ne-Ly$\beta$ line. However, in the CE plasma, the
Ne-Ly$\beta$ line flux must be weaker than the Ne-Ly$\alpha$ line flux
at 1.02 {\rm keV}.  In order to get the line flux ratio of Ne-Ly$\alpha$
to Ly$\beta$, we fit the spectra using the Gaussian lines model with an
absorption.  The absorption fixed at $N_{\rm H} =$7~$\times$
10$^{21}{\rm cm}^{-2}$ derived from the fitting result of the CE plasma
model.  Figure~\ref{fg:f6-3} shows the confidence contour map of the
normalizations of two Gaussian lines (Ne-Ly$\alpha$ and Ly$\beta$) and
indicates that the residual at 1.2 {\rm keV} is significant at the
3$\sigma$ level. The best-fit ratio of Ly$\beta$/Ly$\alpha$ is about
unity that is much larger than the ratio for the CE plasma,
$\sim0.3$. The high Ly$\beta$/Ly$\alpha$ ratio is often obtained from a
plasma in a recombining phase.

From the above consideration, there is a possibility that the residuals
at 0.5, 0.73, 0.87, 1.21 and 1.37 {\rm keV} result from radiative
recombination continuum (RRC) edges of
C$\emissiontype{V\hspace{-.1em}I}$,
O$\emissiontype{V\hspace{-.1em}I\hspace{-.1em}I}$,
O$\emissiontype{V\hspace{-.1em}I\hspace{-.1em}I\hspace{-.1em}I}$,
Ne$\emissiontype{I\hspace{-.1em}X}$, and Ne$\emissiontype{X}$,
respectively.  We therefore introduce non-equilibrium recombining
collisional plasma model ({\tt vrnei}\footnote{See
http://heasarc.gsfc.nasa.gov/xanadu/xspec/manual/XSmodelRnei.html}).
This model is based on the assumption that the plasma changes from the
collisional equilibrium state, with the initial temperature
$k_{\rm{B}}T^{\rm cool}_{\rm init}$, to the non-equilibrium recombining
state, with the temperature $k_{\rm{B}}T^{\rm cool}$. The elemental
abundances except for N, O, Ne and Mg were fixed at the best-fit values
for the colliding wind emission (see $\S$~\ref{subsubsec:globalfitAB}
and table~\ref{tb:t3-2}). The reduced $\chi^{2}$ was significantly
improved. The best-fit results are shown in table~\ref{tb:t3-4}. The
temperature $k_{\rm{B}}T^{\rm cool}$ is as low as $\sim$21 eV.

\begin{table}
\caption{The best-fit values of the elemental abundance ratios of the variable components (warm and hot components) and their absorption components
by number relative to helium.}
\label{tb:t3-2}
\begin{center}
\begin{tabular}{l c}
 \hline\vspace{-3mm}\\
Element &Abundance ratio\footnotemark[$*$]\\
\hline
He&1\footnotemark[$^{f}$]\\
C&0.4 \footnotemark[$^{f}$]\\
N&0\footnotemark[$^{f}$]\\
O&7.2~$\times$ 10$^{-2}$\footnotemark[$^{f}$]\\
Ne&(6.0$\pm 0.2$)~$\times$ 10$^{-3}$\\
Mg&(5.4$\pm 0.3$)~$\times$ 10$^{-4}$\\
Si&5.9~$\times$ 10$^{-4}$\footnotemark[$^{f}$]\\
S&2.4~$\times$ 10$^{-4}$\footnotemark[$^{f}$]\\
Ar&5.3~$\times$ 10$^{-5}$\footnotemark[$^{f}$]\\
Ca&3.2~$\times$ 10$^{-5}$\footnotemark[$^{f}$]\\
Fe&(4.16$^{+0.08}_{-0.14}$)~$\times$ 10$^{-4}$\\
Ni&(1.73$^{+0.03}_{-0.06}$)~$\times$ 10$^{-5}$\\
\hline
\multicolumn{2}{l}{\hbox to 0pt{\parbox{190mm}{\footnotesize
   \par\noindent
 Notes. This result is obtained with the same fitting as \\
that done to obtain the results in table~\ref{tb:t3-1} (observations\\ A
 and B); we simultaneously fitted the FI and BI\\
 spectra of observations A and B with the model\\
 {\tt TBabs} * ( {\tt varabs}$^{\rm warm}$ * {\tt vpshock}$^{\rm warm}$
 $+$\\
 {\tt varabs}$^{\rm hot}$ * {\tt vpshock}$^{\rm hot}$ ). Other parameters
 are\\
 separately shown in table~\ref{tb:t3-1}. 

 \footnotemark[$*$] Abundance ratios by number are expressed
 relative\\ to helium.\\
 Errors and upper limits are at the 90\% confidence level.\\
 The abundance of H was set to zero. We adopted the\\
 abundance ratio C/He $=$ 0.4 by number,  which was taken\\
 from \citet{hillier99}.  The abundance of nickel\\ 
 is linked to that of iron.\\
 \footnotemark[$^{f}$]Value fixed. \\
 \par\noindent
}\hss}}
\end{tabular} 
\end{center}
\end{table}

\begin{figure*}[ht!]
 \begin{center}
  \FigureFile(160mm,160mm){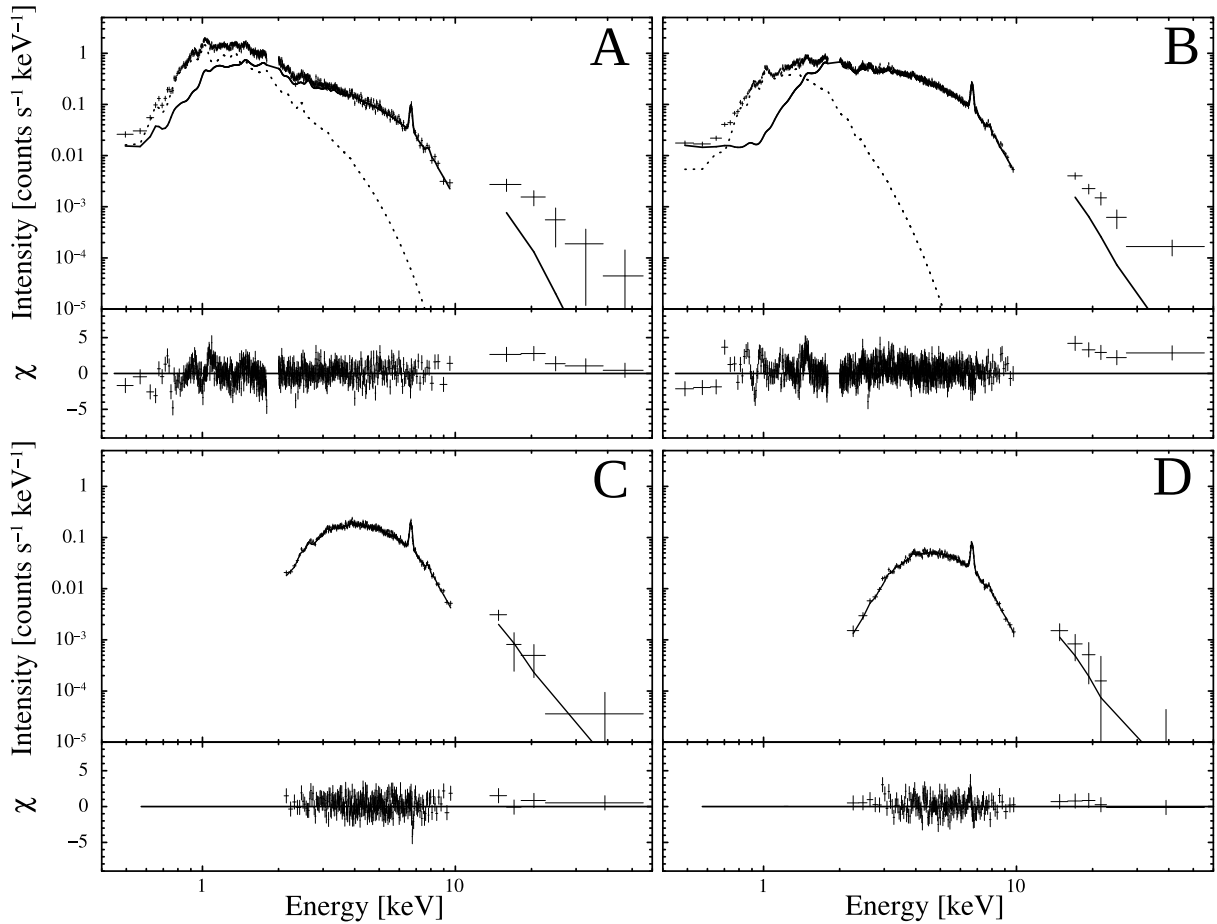}
 \end{center}
\caption{The XIS/FI $+$ HXD/PIN spectra of WR~140 for each observation
 and the XIS fitting models (see table~\ref{tb:t3-2} \&
 \ref{tb:t3-1}). Backgrounds are subtracted. The hot and warm components
 are separately indicated as the solid and dotted lines in the upper
 panels.  The lower panels show the residuals of the data from the
 best-fit model.}  \label{fg:f3-3}
\end{figure*}

\begin{table*}[ht!]
\caption{The best-fit parameters of spectral fitting for the variable components.}
\label{tb:t3-1}
\renewcommand{\footnoterule}{}
\begin{center}
\begin{tabular}{l l c c c c}
\hline\vspace{-3mm}\\
& Observation & A & B & C & D\vspace{1mm}\\
\hline
\vspace{-3mm}\\
& Orbital phase&2.904 &2.989 &2.997 &3.000\vspace{1mm}\\
\hline
\vspace{-3mm}\\
\multicolumn{5}{l}{interstellar absorption}\\
& $N_{\mathrm{H}}$ (10$^{21}$ $\mathrm{cm}^{-2}$)
     &8.51$^{+0.05}_{-0.11}$&$=$ A&$=$ A (fixed)& $=$ A (fixed)\vspace{1mm}\\
\hline
\multicolumn{5}{l}{warm component}\\
 & $N_{\rm{He}}^{\rm warm}$ (10$^{19}$ $\mathrm{cm}^{-2}$)
&--&4.35$^{+0.07}_{-0.10}$&--&--\vspace{1mm}\\
& $k_{\rm{B}}T^{\rm warm}$ ($\mathrm{keV}$)&
0.645$^{+0.014}_{-0.006}$&0.3515$^{+0.0008}_{-0.0006}$&--&--\vspace{1mm}\\
& $\tau_{\rm u}$\footnotemark[$*$](10$^{12}$ s
     cm$^{-3}$)&8.2$^{+2.5}_{-1.2}$&4.1$^{+1.1}_{-0.5}$&--&--\vspace{1mm}\\
& norm \footnotemark[$\dagger$](10$^{-3}$)
&9.97$^{+0.04}_{-0.02}$&14.4$^{+0.3}_{-0.6}$&--&--\vspace{1mm}\\
&$F_{\mathrm{X}}$ \footnotemark[$\ddagger$](10$^{-11}$ $\mathrm{erg}$
     $\mathrm{cm}^{-2}$ $\mathrm{s}^{-1}$) &1.28&0.45&--&--\vspace{1mm}\\
& $L_{\mathrm{X}}$ \footnotemark[$\S$](10$^{34}$ $\mathrm{erg}$ $\mathrm{s}^{-1}$)&2.08 &4.67 &--&-- \vspace{1mm}\\
& $L_{\mathrm{X,bol}}$ \footnotemark[$\|$](10$^{34}$ $\mathrm{erg}$ $\mathrm{s}^{-1}$)&3.84 &10.6 &--&-- \vspace{1mm}\\
\hline
\multicolumn{5}{l}{hot component}\\
 & $N_{\rm{He}}^{\rm hot}$ (10$^{20}$ $\mathrm{cm}^{-2}$)
&$\leq$ 0.007&1.68$^{+0.02}_{-0.03}$&12.6$\pm 0.2$&23.4$\pm 0.5$\vspace{1mm}\\
& $k_{\rm{B}}T^{\rm hot}$ ($\mathrm{keV}$)&
3.29$^{+0.06}_{-0.05}$&3.48$^{+0.04}_{-0.03}$&2.97$^{+0.07}_{-0.05}$&3.03$^{+0.08}_{-0.11}$\vspace{1mm}\\
& $\tau_{\rm u}$\footnotemark[$*$](10$^{12}$ s
     cm$^{-3}$)&8.2\footnotemark[$\#$]&4.1\footnotemark[$\#$]&2.7$^{+1.9}_{-0.8}$&$\geq$
		     2.5\vspace{1mm}\\
& norm \footnotemark[$\dagger$](10$^{-2}$)&1.27$\pm 0.01$&3.45$^{+0.01}_{-0.02}$&3.56$^{+0.08}_{-0.10}$&1.89$^{+0.10}_{-0.08}$
\vspace{1mm}\\
&$F_{\mathrm{X}}$ \footnotemark[$\ddagger$](10$^{-11}$ $\mathrm{erg}$
     $\mathrm{cm}^{-2}$ $\mathrm{s}^{-1}$) &2.85&5.63&2.58&1.01\vspace{1mm}\\
& $L_{\mathrm{X}}$ \footnotemark[$\S$](10$^{34}$ $\mathrm{erg}$ $\mathrm{s}^{-1}$
     )&1.67 &5.10 &5.78 &2.51 \vspace{1mm}\\
& $L_{\mathrm{X,bol}}$ \footnotemark[$\|$](10$^{34}$ $\mathrm{erg}$ $\mathrm{s}^{-1}$
     )&2.59 &9.08 &11.7 &4.13 \vspace{1mm}\\
\hline\vspace{1mm}
 & $\chi^{2}$/dof &\multicolumn{2}{c}{3744/2330} &556/472&297/253\\
\hline
 \multicolumn{5}{l}{\hbox to 0pt{\parbox{190mm}{\footnotesize
   \par\noindent
  Notes. 
For observations A and B, we adopted the model {\tt TBabs} * ( {\tt
 varabs}$^{\rm warm}$ * {\tt vpshock}$^{\rm warm}$ $+$ \\{\tt
 varabs}$^{\rm hot}$ * {\tt vpshock}$^{\rm hot}$ ). In observation~A, we
 removed {\tt varabs}$^{\rm warm}$ in order to determine the
 interstellar\\ absorption for WR~140. On the other hand, for observations
 C and D, we adopted the model {\tt TBabs} * {\tt varabs}$^{\rm hot}$
 * \\{\tt vpshock}$^{\rm  hot}$. Errors and upper limits are at the 90\%
 confidence level.
 The elemental abundance ratios of the variable\\
 components (warm and hot components) and the absorption components
are separately shown in table~\ref{tb:t3-2}.\\ 
 \footnotemark[$*$] Ionization timescale\\
\footnotemark[$\dagger$] Normalization constant defined as ${\it E.M.} \times
 10^{-14}(4 \pi {\it d}^{2})^{-1}$, where ${\it E.M.}$ is the emission
 measure in $\rm{cm}^{-3}$ \\ and {\it d} is the distance in {\rm cm}.\\
 \footnotemark[$\ddagger$] The absorbed flux (0.5--10 {\rm keV})\\
 \footnotemark[$\S$] The absorption-corrected luminosity (0.5--50
 {\rm keV}), which we calculated by assuming a distance of 1.67 {\rm kpc}.\\
   \footnotemark[$\|$] The intrinsic bolometric luminosity\\
 \footnotemark[$\#$]Ionization timescale of the hot component is
 tied to that of warm component.\\
   \par\noindent
   }\hss}}
\par\noindent
\end{tabular}
\end{center}
\end{table*}

\subsection{Variable component}
\label{subsec:variable_comp} 
The emission from variable component constitutes a considerable fraction
of the X-ray emission from WR~140 (see figure~\ref{fg:f3-2}).  For the
spectral analysis, we then adopted the source and background regions
used in $\S$~\ref{subsec:spectral_chara}.  As for observations C and D,
we ignored the soft band below $\sim$2~{\rm keV} because there is the cool
component (see figure~\ref{fg:f3-3}).

\subsubsection{Spectral fitting}
\label{subsubsec:globalfitAB}

In this subsubsection, we made the FI and BI spectra of observations A
and B in the 0.5--10~{\rm keV} band. We executed simultaneous fitting of
the FI and BI spectra of observations A and B.  We first tried to fit
these spectra using a one-temperature plasma emission model with either
a single absorption component or a combination of absorption components
with two different optical depths.  We next tried a two-temperature
plasma emission model with a single common absorption.  Both the
attempts did not produce statistically acceptable solutions.  This is
perhaps because one-temperature plasma models do not reproduce emission
lines in the soft band around 1~{\rm keV} and iron K line at
$\sim$6.7~{\rm keV} simultaneously, and those spectra, in particular in
observation B, are too flat to be reproduced with a single absorption
component.

We therefore fit the XIS spectra with a two-temperature (warm and hot)
plasma model with independent absorption components ({\tt
varabs})\footnote{We found that the line shift, which we will discuss in
$\S$~\ref{subsubsec:FeKprofile}, had little effect on the overall fit,
so that we did not change the red-shift parameter in the spectral
model.}. For the emission components, we first used the CE plasma
emission code, {\tt vAPEC} \citep{smith01} in which we can allow to vary
each metal abundance. Next, we used the plane-parallel shock code, {\tt
vpshock}.  In both of the fits, we fixed the C/He abundance ratio at 0.4
by number \citep{hillier99} and the other elements except for Ne, Mg, Fe
and Ni at the best-fit abundances in \citet{pollock05}.  The latter fit
gave slightly better reduced $\chi^{2}$ values (2Tvapec: $\chi^{2}$/dof
$\sim$1.71, 2Tvpshock: $\chi^{2}$/dof $\sim$1.61). The top panels in
figure~\ref{fg:f3-3} show the best-fit 2Tvpshock models and the spectra.
We use the {\tt vpshock} model in the following analysis.

The X-ray emission should be absorbed both in the W-R wind and ISM. We
therefore assume two absorption components.  One is the absorption
component ({\tt varabs}) for the W-R wind, whose elemental abundances
are tied to those of the plasma ({\tt vpshock}) model. Another is an
interstellar absorption component, with elemental abundances fixed at
ISM abundances ({\tt TBabs}: \cite{wilms00}). We also assumed that the
warm component in observation A suffers only interstellar absorption.

We tied elemental abundances of the both plasmas for observations A and
B, since the elemental abundance of the plasma should not change between
observations A and B. We also tied the ionization parameters $\tau_{\rm
u}$ for both hot and warm components.  The best-fit results are given in
tables~\ref{tb:t3-2} and \ref{tb:t3-1}. The best-fit model for the
spectrum of observation A gives an absorption column of $N_{\rm H}
=$8.51~$\times$ 10$^{21}{\rm cm}^{-2}$. The observation A is correspond
to orbital phase $=2.904$, in the definition of \citet{monnier11}. The
column density is close to that obtained with an earlier observation
with {\it Chandra} ($N_{\rm H} =$8.0~$\times$ 10$^{21}{\rm cm}^{-2}$ at
orbital phase $=1.986$; \cite{pollock05}).

The interstellar extinction toward WR~140 has been measured at the
optical wavelength as $A_{V} =$ 2.95 {\rm mag} \citep{morris93}. Based
on the $A_{V}$, the expected ISM absorption column ($N_{\rm H}$) is
either 6.5~$\times$ 10$^{21}$ (using the correlation of
\cite{gorenstein75}) or 4.6~$\times$ 10$^{21}{\rm cm}^{-2}$(using the
correlation of \cite{vuong03}). In these correlations, we assumed that
the gas-to-dust ratio of WR~140 is the same as that in ISM.  The
observed absorption at the phase A differs from the expected absorptions
only by a factor of two. The absorption appearing in the data of
observation A may be mainly made by the ISM toward WR~140. For
simplicity, we regard the X-ray absorption in the spectrum of
observation A as the interstellar absorption hereafter. No conclusion in
this paper changes with this simplification.

We fit the spectra of observations C and D independently. We executed
simultaneous fitting of FI and BI spectra. Since the soft band flux
below $\sim$4~{\rm keV} is strongly suppressed by circumstellar
absorption, we did not include the warm component in the fits to either
the spectrum of observation C or that of observation D. The elemental
abundance were fixed at the best-fit values of observations A and B (see
table~\ref{tb:t3-2}). The bottom panels in figure~\ref{fg:f3-3} and
table~\ref{tb:t3-1} show the best-fit results.

\begin{table*}[ht!]
\caption{Results of spectral fitting in the 5--9 {\rm keV} band.}
 \label{tb:t3-6}
\begin{center}
\begin{tabular}{l lcccc}
\hline
\vspace{-3mm}\\
 Obs.& I.D.&A & B & C & D \vspace{1mm}\\
\hline
\vspace{-3mm}\\
\multicolumn{6}{l}{Power-law}\vspace{1mm}\\
$\Gamma$&&3.0$\pm 0.2$&2.98$^{+0.07}_{-0.06}$&3.21$\pm 0.08$&2.98$^{+0.03}_{-0.05}$\vspace{1mm}\\
norm\footnotemark[$*$]&&0.04$\pm 0.01$&0.12$^{+0.02}_{-0.01}$&0.17$^{+0.03}_{-0.02}$&0.06$\pm 0.01$\vspace{1mm}\\
\multicolumn{6}{l}{Gaussian}\vspace{1mm}\\
$E_{1}$\footnotemark[$\dagger$] ($\mathrm{eV}$)&Fe$\emissiontype{XXV}$
K$\alpha$&6697$\pm 9$ $\pm
     7$&6678$^{+3}_{-4}$ $\pm 7$&6668$\pm 4$ $\pm 7$&6661$^{+4}_{-5}$ $\pm 7$\vspace{1mm}\\
{\it v}\footnotemark[$\ddagger$] (10$^{2}~{\rm
 km~s^{-1}}$)&&$-$8.6$^{+4.2}_{-4.5}$ $\pm 3.1$&$-$0.9$^{+1.8}_{-1.4}$
	 $\pm 3.1$ &$+$4.5$^{+2.5}_{-2.3}$ $\pm
	     3.1$&$+$$7.7^{+3.1}_{-2.3}$ $\pm 3.1$\\
$\sigma_{1}$ ($\mathrm{eV}$)&&$<$5&$<$19&$<$15&$<$31\vspace{1mm}\\
$F_{1}$\footnotemark[$\S$] &&5.8$\pm 0.6$&18.1$\pm 0.6$&16.1$\pm 0.6$&9.4$\pm 0.5$\vspace{1mm}\\
$E_{2}$ ($\mathrm{eV}$)&Fe$\emissiontype{XXV\hspace{-.1em}I}$
Ly$\alpha$&6973\footnotemark[$^{f}$]&6973\footnotemark[$^{f}$]&6973\footnotemark[$^{f}$]&6973\footnotemark[$^{f}$]\vspace{1mm}\\
$F_{2}$\footnotemark[$\S$]&&$<$0.5&1.0$\pm 0.4$&0.51$\pm 0.4$&$<$0.2\vspace{1mm}\\
$E_{3}$ ($\mathrm{eV}$)&Ni$\emissiontype{XXV\hspace{-.1em}I\hspace{-.1em}I}$
 K$\alpha$&7797\footnotemark[$^{f}$]&7797\footnotemark[$^{f}$]&7797\footnotemark[$^{f}$]&7797\footnotemark[$^{f}$]\vspace{1mm}\\
$F_{3}$\footnotemark[$\S$] &&$<$0.9&1.3$\pm 0.6$&1.1$\pm 0.7$&0.9$^{+0.3}_{-0.4}$\vspace{1mm}\\
$E_{4}$ ($\mathrm{eV}$)&Fe$\emissiontype{XXV}$ K$\beta$&7897\footnotemark[$^{f}$]&7897\footnotemark[$^{f}$]&7897\footnotemark[$^{f}$]&7897\footnotemark[$^{f}$]\vspace{1mm}\\
$F_{4}$\footnotemark[$\S$]&&$<$0.8&$<$0.9&0.8$\pm 0.7$&$<$0.3\vspace{1mm}\\
\hline
\vspace{-3mm}\\
 $\chi^{2}$/dof&&128/96&278/238&192/158&151/159\\
\hline
Mn K$\alpha$\footnotemark[$\|$] ($\mathrm{eV}$)&&5904$^{+5}_{-3}$&5889$^{+4}_{-2}$&5896$\pm 2$&5898$^{+1}_{-3}$\\
\hline
\multicolumn{5}{@{}l@{}}{\hbox to 0pt{\parbox{190mm}{\footnotesize
   Notes. Errors and upper limits are at the 90\% confidence
 level. Subscript Numbers show \\1$=$Fe$\emissiontype{XXV}$
K$\alpha$, 2$=$Fe$\emissiontype{XXV\hspace{-.1em}I}$
Ly$\alpha$, 3$=$Ni$\emissiontype{XXV\hspace{-.1em}I\hspace{-.1em}I}$
 K$\alpha$ and 4$=$Fe$\emissiontype{XXV}$ K$\beta$.\\
   \footnotemark[$*$] Normalization constant, defined as the flux
 density at 1 keV in units of photons $\mathrm{keV}^{-1}~\mathrm{cm}^{-2}~\mathrm{s}^{-1}$.\\
   \footnotemark[$\dagger$] The first and second errors show a statistical error and a
 systematic error, respectively.\\
   \footnotemark[$\ddagger$] Line radial velocity. Relative velocities
 are based on the line center energy of  Fe$\emissiontype{XXV}$
 K$\alpha$$\sim$6,676~{\rm eV} in the rest frame.\\
   \footnotemark[$\S$] Line flux in 10$^{-5}$ photon $\mathrm{cm}^{-2}~\mathrm{s}^{-1}$.\\
   \footnotemark[$^{f}$] Line center values were fixed.\\
\footnotemark[$\|$] The center energy of Mn K$\alpha$ line from the
 $^{55}$Fe calibration source.\\  
   \par\noindent
 }\hss}}
\par\noindent
\end{tabular} 
\end{center}
\end{table*}

\begin{figure}[ht!]
 \begin{center}
 \FigureFile(85mm,85mm){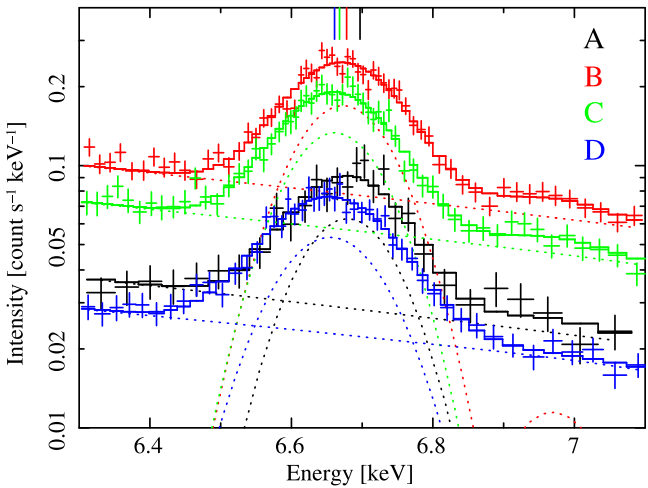}
 \end{center}
\caption{The zoomed XIS/FI spectra of WR~140 for each observation and
 the fitting models (see table~\ref{tb:t3-6}.) The black, red, green and
 blue show the observation A, B, C and D, respectively. The dotted lines
 show the power-law and Gaussian components. The solid lines show their
 summation. Vertical lines indicate the central energies of the
 Fe$\emissiontype{XXV}$ K$\alpha$ line for observations A, B, C and D.}
 \label{fg:Fe}
\end{figure}

\subsubsection{Fe K-line Profile}
\label{subsubsec:FeKprofile}

The XIS spectra showed emission lines of Fe$\emissiontype{XXV}$
K$\alpha$ ($\sim$6.7 {\rm keV}), Fe$\emissiontype{XXV\hspace{-.1em}I}$
Ly$\alpha$ ($\sim$6.9 {\rm keV}),
Ni$\emissiontype{XXV\hspace{-.1em}I\hspace{-.1em}I}$ K$\alpha$
($\sim$7.8 {\rm keV}) and Fe$\emissiontype{XXV}$ K$\beta$ ($\sim$7.9
{\rm keV}).  In order to investigate the variation of their line
profiles, we measured the center energies, widths and fluxes of these
lines using the FI data which has better gain calibration and higher
sensitivity in the hard energy band. We fitted the spectra in the
5--9~{\rm keV} band with a continuum model and four Gaussian models for
the lines. For the continuum emission, we tried the two different
models, thermal bremsstrahlung and power-law.  Both fits gave consistent
results. In fact, the best-fit parameters of the lines for two different
continuum models are found to be within the margin of error. We
therefore show only the best-fit parameters of a power-law model for the
continuum and four Gaussian models for the lines. We fixed the
line-center energies, except for that of Fe$\emissiontype{XXV}$
K$\alpha$ line, at the rest frame values, due to poor photon
statistics. Figure~\ref{fg:Fe} and table~\ref{tb:t3-6} show the best-fit
results. The advantage of Suzaku-XIS is that the energy-scale is
calibrated very well with the accuracy of roughly 0.1\% (e.g.,
\cite{koyama07b}; \cite{ota07}; \cite{ozawa09};
\cite{tamura11}). Therefore, we adopted the systematic errors of 0.1\%
(7 {\rm eV}) in our result, as are shown in table~\ref{tb:t3-6}.

\begin{figure}[ht!]
 \begin{center}
 \FigureFile(85mm,85mm){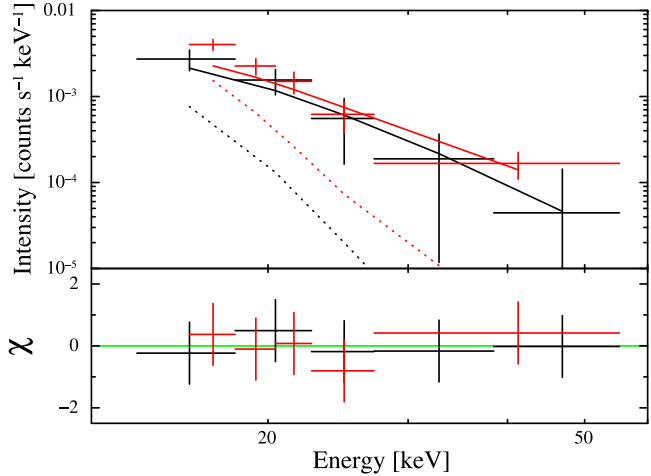}
 \end{center}
\caption{The HXD/PIN spectra of WR~140 for observations A and B and the
 best-fit models. The black and red show observations A and B,
 respectively. The solid and dotted lines in the upper panel show the
 power-law and hot components, respectively. Lower panel shows the
 ratios of the data to the best-fit model. See also
 table~\ref{tb:t3-5}.}  \label{fg:pin}
\end{figure}

As the binary system approached periastron, the central energy of
Fe$\emissiontype{XXV}$ K$\alpha$ line gradually shifted from
6,697$\pm9$~{\rm eV} (observation A) to 6,661$^{+4}_{-5}$~{\rm eV}
(observation D). In order to measure the shift of the central energy of
the Fe$\emissiontype{XXV}$ K$\alpha$ line, we determined the line center
energy in the rest frame for the CE plasma model
($k_{\rm{B}}T\sim$3~{\rm keV}).  Here, we used the {\tt fakeit} of {\tt
Xspec} to take account of the influences of not only dominant iron lines
but also other weak lines derived from other elements. The line center
energy was $\sim$6,676~{\rm eV}.

We have to pay attention to the fact that the line center energy could
shift with variations of the flux ratios between the resonance,
intercombination and forbidden lines, which are caused by
non-equilibrium ionization. The ionization timescale $\tau_{\rm u}$
changed by a factor of three from observation A to D (see
table~\ref{tb:t3-1}).  In order to check the influences of the
difference of the ionization timescale, we estimated the center energies
based on the {\tt fakeit} with the non-equilibrium ionization models
(table~\ref{tb:t3-1}). The center energies for non-equilibrium
ionization models had no difference and were exactly similar to that for
the CE plasma model in the result. Therefore, we regarded the observed
shifts in the line center energy as the Doppler shift.

The radial velocity changed from $-$860 to $+$770~{\rm km s$^{-1}$} from
observation A to that of D. The radial velocities in table~\ref{tb:t3-6}
were calculated by adopting the rest frame energy as 6,676~{\rm eV}.
This is consistent with the earlier result of ${\it Chandra}$
observations obtained by \citet{pollock05} (orbital phase $=$ 1.986 and
2.032). They reported that Fe$\emissiontype{XXV}$ line was shifted from
$-$600 {\rm km s$^{-1}$} to$+$648$\pm 259~{\rm km~s^{-1}}$ during the
periastron passage. Such a variation was also seen in
He$\emissiontype{I}$ line in the near-infrared band during the
periastron passage in 2001 \citep{varricatt04}.
In addition, \citet{marchenko03} reported similar shifts of
C$\emissiontype{III}$ and He$\emissiontype{I}$ lines at optical
wavelength, which are thought to be emitted from wind-wind collision
layer.  Our observed variation of line center energy is consistent with
such Doppler motions. This indicates that the emission in all
wavelengths comes from a common shock cone.

\subsection{Hard excess above 10 keV}

\begin{table}[ht!]
\caption{Results of spectral fitting in the 15--50 {\rm keV} band.}
\label{tb:t3-5}
\renewcommand{\footnoterule}{}
\begin{center}
\begin{tabular}{l l c c}
\hline\vspace{-3mm}\\
& Observation & A & B \vspace{1mm}\\
\hline
\vspace{-3mm}\\
& Orbital phase&2.904 &2.989\vspace{1mm}\\
\hline
\vspace{-3mm}\\
\multicolumn{4}{l}{power-law component}\\
 &$\Gamma$ &1.7$^{+1.1}_{-0.8}$& = A \vspace{1mm}\\
 &norm \footnotemark[$*$] (10$^{-3}$)&1.5$^{+36.8}_{-1.4}$&1.8$^{+51.0}_{-1.7}$\vspace{1mm}\\
&$F_{\mathrm{X}}$ \footnotemark[$\dagger$](10$^{-12}$ $\mathrm{erg}$
     $\mathrm{cm}^{-2}$ $\mathrm{s}^{-1}$) &7.3&9.0\vspace{1mm}\\
 & $L_{\mathrm{X}}$ \footnotemark[$\ddagger$](10$^{33}$ $\mathrm{erg}$ $\mathrm{s}^{-1}$
     )&6.2 &7.6\vspace{1mm}\\
\hline\vspace{1mm}
 & $\chi^{2}$/dof &\multicolumn{2}{c}{1.34/7}\\
\hline
 \multicolumn{4}{l}{\hbox to 0pt{\parbox{190mm}{\footnotesize
   \par\noindent
  Notes. 
  We adopted the model {\tt TBabs} * ( ({\tt varabs}$^{\rm warm}$ * \\
 {\tt vpshock}$^{\rm warm}$) $+$ {\tt varabs}$^{\rm hot}$ * ({\tt
 vpshock}$^{\rm hot}$ $+$ {\tt powerlaw})).\\ We fixed the warm and hot
 components at the best-fit values (see\\
 tables~\ref{tb:t3-2} and ~\ref{tb:t3-1}). Errors and upper limits are at the 90\% confidence level.\\ 
 \footnotemark[$*$] Normalization constant of the power-law model
 defined as photons\\ {\rm keV}$^{-1}~{\rm cm}^{-2}~{\rm s}^{-1}$ at 1
 {\rm keV}.\\
 \footnotemark[$\dagger$] The absorbed flux (15--50 {\rm keV})\\
 \footnotemark[$\ddagger$] The absorption-corrected luminosity (0.5--50
 {\rm keV}) was calculated\\ assuming a distance of 1.67 {\rm kpc}.\\
   \par\noindent
   }\hss}}
\par\noindent
\end{tabular}
\end{center}
\end{table}

Figure~\ref{fg:f3-3} shows the XIS and HXD/PIN spectra and the XIS
best-fit models of the variable component\footnote{Since the {\tt
vpshock} model does not cover the energy band above 10~{\rm keV}, we used the
{\tt vAPEC} model to estimate the thermal contribution in the HXD energy
range.}. As explained in $\S$~\ref{subsec:spectral_chara}, the combined
spectra of NXB and CXB were used as the HXD/PIN background spectra. In
both A and B observations, we found the hard-tail excess emission in the
HXD/PIN band. The excess count rates after deduction of the emission
from variable component at observations A and B are 0.022 and 0.021
counts per second in the 15--50~{\rm keV} band, respectively. On the other
hand, in observations C and D, there are no significant excess
emission in the HXD/PIN band spectra.

Hereafter, we fit the excesses from the warm and hot components in
observations A and B, respectively.  We introduce a power-law model for
the hard band excess. We simultaneously fit HXD/PIN spectra of
observations A and B, tying the power-law index, since the HXD/PIN
spectra are limited in photon statistics. Since the spectra do not
constrain the soft band shape of the hard band excess, we assumed the
same absorption to the excess component as that to the hot component in
each phase. The best-fit model yields a power-law index
$\Gamma=1.7$. However, the hard band excess could be plausibly
reproduced with a bremsstrahlung model with $k_{\rm{B}}T$ $>$10~{\rm
keV}, too.  The results are shown in figure~\ref{fg:pin} and
table~\ref{tb:t3-5}.

We have to consider the contamination from the other hard X-ray source
IGR J20216+4359, because this source is located in the field of view of
HXD/PIN. We discuss this contamination in detail in
$\S$~\ref{subsec:hard-tail}.

   \begin{figure}[ht!]
   \begin{center}
    \FigureFile(80mm,80mm){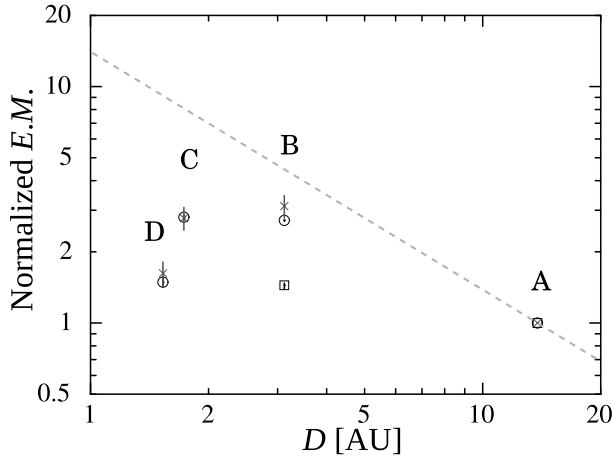}
   \end{center}
   \caption{The variation of the normalized emission measure and
   normalized line flux ($\Box$ warm component, $\circ$ hot component,
   $\times$ Fe line) with stellar separation. The dashed line shows an
   inverse proportion to the distance between the two stars.}
   \label{fg:f3-5}
  \end{figure}

 \begin{figure}[ht!]
    \begin{center}
     \FigureFile(85mm,85mm){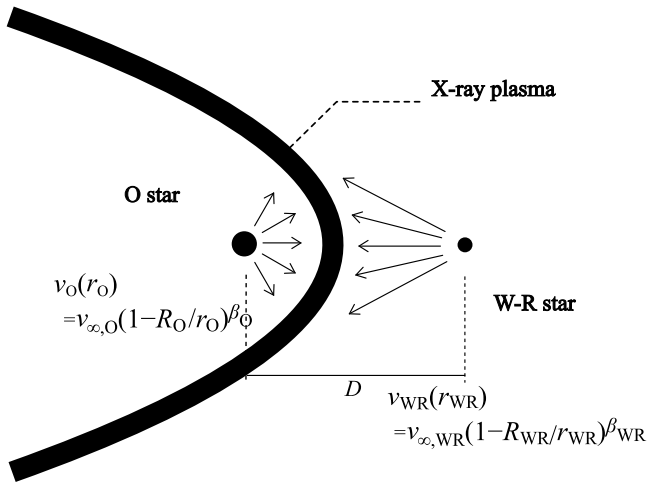}
    \end{center}    
    \caption{A cartoon of a colliding-wind binary. $\it{D}$, $\it{r}_{\rm
    O}$ and $\it{r}_{\rm WR}$ show binary separation, the distance from
    the O star to the colliding-wind region and the distance from the
    W-R star to the colliding wind region, respectively.}
   \label{fg:f4-2}
   \end{figure}

\section{Discussion}

The warm, hot and hard band excess components vary with orbital phase.
The cool component is detected only from observations C and D.  The cool
component appears nearly constant.

\subsection{The nature of the cool component}

The cool component was discovered for the first time with this
observation. The simple analysis of one-temperature CE plasma emission
for the cool component failed to reproduce the observed spectra with the
residuals at 1.21~{\rm keV} and other energies remaining in the spectral
fitting.  The energy of the residuals coincides fairly well with the
free-bound absorption features of the highly ionized ions of O, Ne and
Mg elements. The fit is then significantly improved with the recombining
collisional plasma model, which introduces the radiative recombination
continua (RRC).

   \begin{figure}[h]
   \begin{center}
    \FigureFile(85mm,120mm){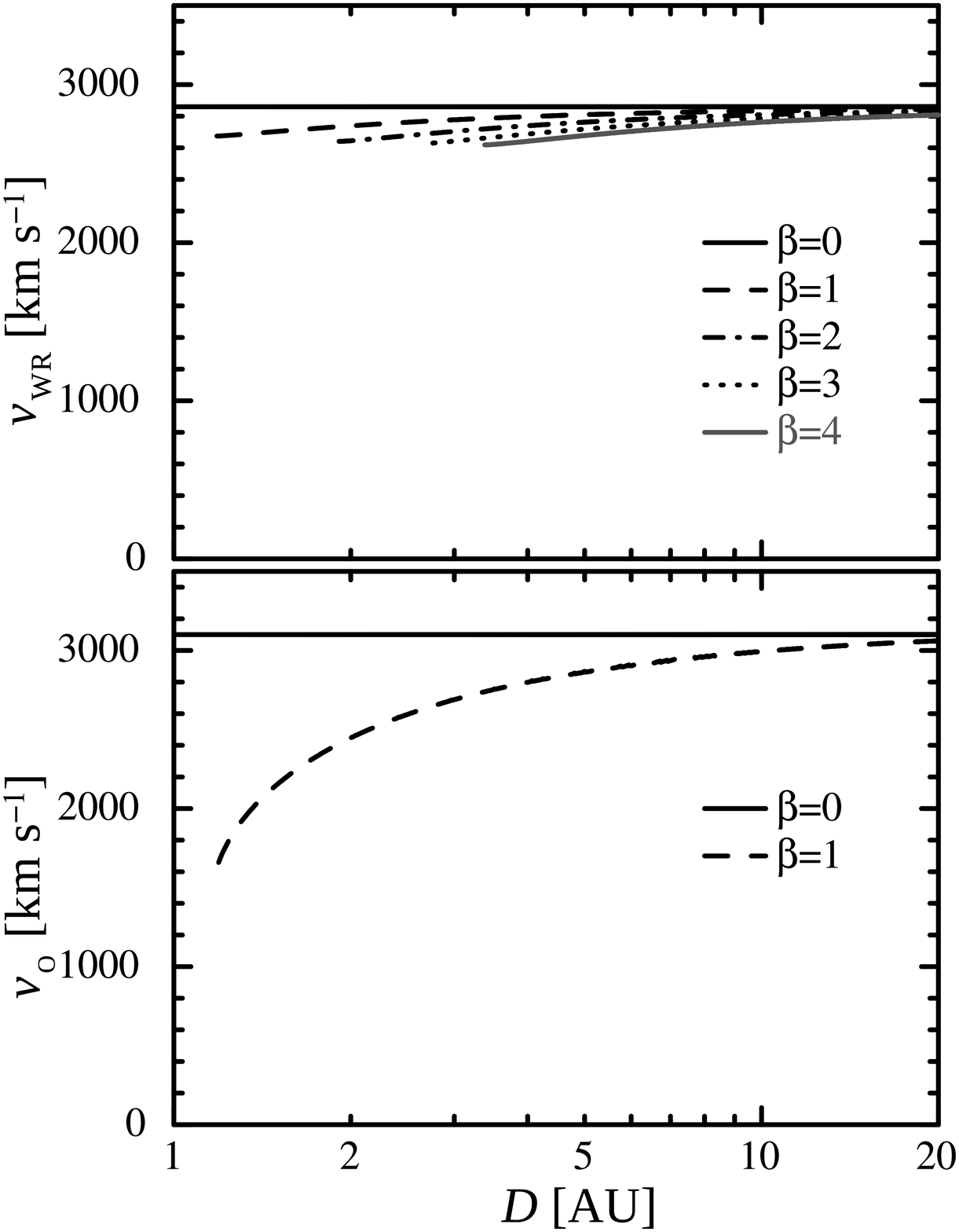}
   \end{center}
   \caption{The binary separation vs. wind velocity at the stagnation
   point. Top: W-R star. Bottom: O star.}  
   \label{fg:f4-3}
  \end{figure}

The Ly$\beta$/Ly$\alpha$ line flux ratio of Ne in the cool component is
also found to be significantly higher than that expected for the CE
plasma (figure~\ref{fg:f6-3}). The high flux ratio can be understood
with a transition process from a recombination of free electrons.  Both
the high Ly$\beta$/Ly$\alpha$ ratio and the RRC features indicate that
the cool component is interpreted as a collisional plasma in a
recombining phase.

The small absorption column of an order of $10^{21}$ H cm$^{-2}$ (see
table~\ref{tb:t3-4}) constrains the location of the cool component
(figure~\ref{fg:3-4}).  The spectra of the hot components are heavily
absorbed below 1.6 keV at observations C and D.  The large
difference in the absorptions between both components indicates that the
cool component is not deeply embedded in the wind-wind collision layer
at observations C and D.  Moreover, the column density for the cool
component is similar to or is rather smaller than that for the warm and hot
components at observation A, in which the distance of colliding wind
region to the W-R star is about 10 AU.  This suggests that the cool plasma
is located far away from the W-R star by 10 AU or more.

In conclusion, the cool component that we discovered can be interpreted
as a collisional plasma in a recombining phase.  A potential scenario is
as follows.  The plasma was heated in the past by a wind-wind collision
shock. The plasma then escaped from the dense wind region, and now, is
recombining electrons and emitting the radiative recombination continuum
along with the collisional plasma emission.

 \begin{figure*}[ht!]
   \begin{center}
    \FigureFile(170mm,170mm){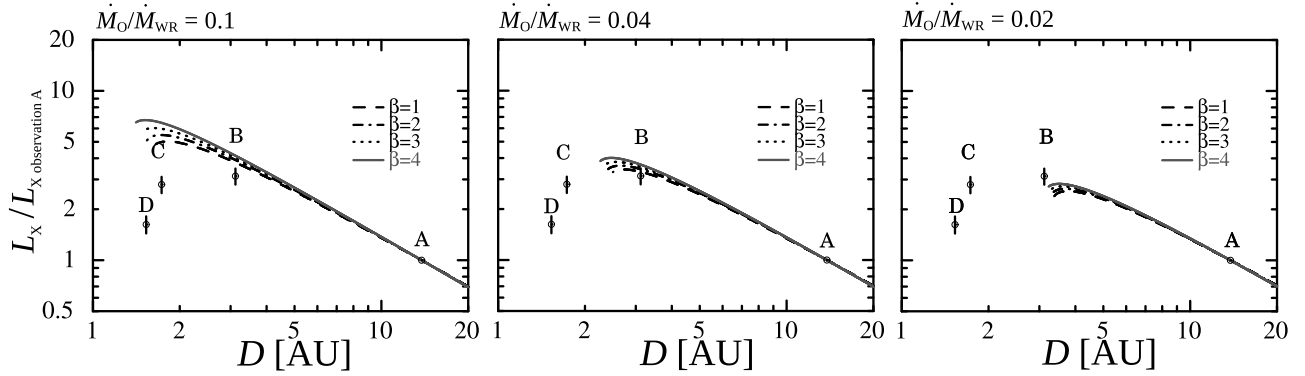}
   \end{center}
   \caption{Binary separation vs. normalized X-ray luminosity. The
  luminosity is normalized at the separation of observation A. $\dot
  M_{\rm O}/\dot M_{\rm WR}$ in the left, center and bottom panels are
  0.1, 0.04 and 0.02, respectively. The $\beta_{\rm O}$ for the O star
  is fixed to be unity. The lines show the equation (\ref{eq1}).}
  \label{fg:f4-6}
  \end{figure*}

The initial temperature $k_{\rm{B}}T^{\rm cool}_{\rm init}$ of the cool
component was derived from the fitting results with the recombining
collisional plasma model (table~\ref{tb:t3-4}).  The range of
$k_{\rm{B}}T^{\rm cool}_{\rm init}$ contains the temperature of the
warm and hot components, which were formed by the wind-wind collision.
On the one hand, the plasma temperature $k_{\rm{B}}T^{\rm cool}$ is now
0.021~{\rm keV} (table~\ref{tb:t3-4}).  This temperature of the cool component
is between the temperature of the wind-wind collision plasma and that of
the dust detected in the infrared wavelengths \citep{williams09}.  The
observed cool component might be emitted from the area in a {\it
transition} phase, where the hot gas, compressed by the wind-wind
collision, is cooling down and is forming dust.

The cooling of the shock heated plasma in wind-wind collision layer
should be common in all the W-R binaries. If so, is the recombining
plasma with less absorption detected commonly in them?  The RRC
structures in the X-ray spectra were reported from the two bright
colliding wind binaries $\gamma^2$ Velorum (WC8 $+$ O7.5:
\cite{schild04}) and $\theta$ Muscae (WC5 $+$ O6 $+$ O9.5:
\cite{sugawara08}). At least, the emission component with the RRC
structure of $\gamma^2$ Velorum is less absorbed. $\gamma^2$ Velorum and
$\theta$ Muscae are both WC-subtype W-R binaries. On the other hand, no
detection of the RRC structure was reported from the other subtype WN.
Systematic and deep search of the structure from W-R binaries is
critical to answer the question.

 \subsection{Flux variation of the hot component}
\label{subsec:flux} 

From table~\ref{tb:t3-1}, we can see that the emission measure of the
 hot component varied. Generally, in colliding wind binaries, the
 emission measure in colliding wind region is expected to vary along
 with inversely proportional relation with the binary separation {\it D}
 \citep{usov92}. In order to check whether WR~140 varied with this
 relation, we plotted the emission measure of the hot component in
 figure~\ref{fg:f3-5}. The emission measure of the warm component and
 Fe$\emissiontype{XXV}$ K$\alpha$ line flux are also plotted in the same
 figure.  As the phase approaches toward periastron, to observation C
 and then to D, the emission measure of the hot component deviated from
 the expected 1/{\it D} law. No numerical simulation ever published
 predicts a rapid flux decrease at periastron, although some theorists
 simulated the variation of the X-ray luminosity for the colliding-wind
 binary (e.g., \cite{luo90}; \cite{stevens92}; \cite{myasnikov93}).

 We therefore took account of not only the separation {\it D}, but also
 another parameter. According to the equation (10) in \citet{stevens92},
 the intrinsic X-ray luminosity of the colliding wind zone can be
 written as
  \begin{eqnarray}                                                    
   {L_{\rm X}} \propto {\it                                           
    D}^{-1} (1+{\it A})/{\it A}^{4},                                  
    \label{eq1}                                                       
  \end{eqnarray}                                                      
  where the wind momentum flux ratio, ${\it A} = \left((\dot M_{\rm
  WR}~v_{\rm WR}(r_{\rm WR}))/(\dot M_{\rm O}~v_{\rm O}(r_{\rm
  O}))\right)^{1/2}$. Here, $\it{r}_{\rm O}$ and $\it{r}_{\rm WR}$ show
  the distances to the shock region from the O star and from the W-R
  star, respectively. The geometry is sketched in figure~\ref{fg:f4-2}.
  Note that {\it A} is not constant and changes depending on not only
  the binary separation {\it D} but also the wind acceleration parameter
  $\beta$ and the ratio of mass-loss rates.

  At periastron, the momentum flux of the O-type wind should become
  significantly smaller, since its wind has less space to reach terminal
  velocity before entering the shock region, while the W-R wind should
  be less affected. In figure~\ref{fg:f4-3}, we show the maximum
  pre-shock stellar wind velocities of the W-R and O stars at a given
  binary separation. We adopted a simple beta law for the wind
  acceleration: $v(r)=v_{\infty}(1-R/r)^{\beta}$. Here, $v_{\infty}$ and
  $R$ are the terminal wind velocity and the stellar radius,
  respectively. We used the value of $v_{\infty,\rm{WR}}=$~2860~${\rm
  km~s}^{-1}$ \citep{williams90}, $v_{\infty,\rm{O}}=$~3100~${\rm
  km~s}^{-1}$ \citep{setia01}, $R_{\rm WR}=~2~{\rm R}_{\odot}$ and
  $R_{\rm O} = 26~{\rm R}_{\odot}$ (cf. \cite{williams09}).  Many
  studies (e.g., \cite{puls96}; \cite{lepine99}) indicate that the
  $\beta$ value of O star winds should be near 1, while W-R stars can
  have larger values $\beta >$1. We therefore fixed the $\beta$ value of
  O star wind to unity in the following simulations. The actual $\beta$
  value for the W-R star is not so important since the stagnation point
  is far from the W-R star. Near the shock region, the W-R wind reaches
  its terminal velocity for any assumed $\beta$, while the O-star wind
  lags far behind. As a result, the wind momentum flux ratio, {\it A},
  rapidly becomes large near periastron.  This means that the stagnation
  point approaches the O star.

  In figure~\ref{fg:f4-6}, we plot the normalized X-ray luminosity as a
  function of the binary separation $D$.  The luminosity is normalized
  at the separation of observation A ($D \sim$ 14 {\rm AU}). In this
  figure, we adopted the X-ray luminosity ratio of the hot
  components. The bolometric X-ray luminosity ratio of the warm
  component to the hot one is about one in observations A and B
  (table~\ref{tb:t3-1}). We assumed that the ratios in observations C
  and D remain unchanged. The observed luminosity peak appears near
  observation B.  The luminosity drops at observations C and D and
  indicates the wind momentum ratio cannot be constant; $\beta \neq 0$.

\begin{figure}[ht!]
 \begin{center}
  \FigureFile(80mm,80mm){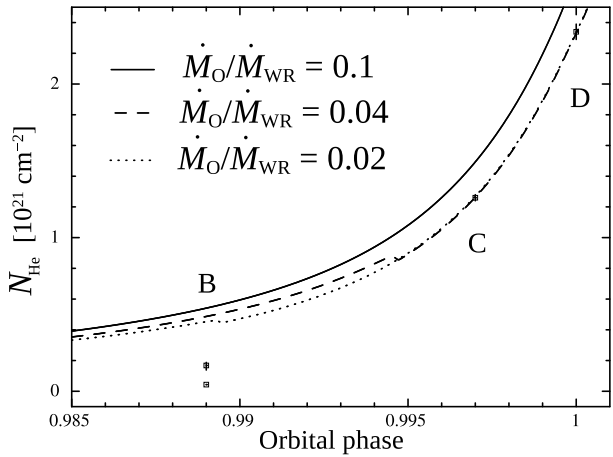}
 \end{center}
 \caption{The helium column density for the warm and hot components
 plotted against the orbital phase. Observed absorptions are plotted by
 a bar and the calculated absorption by solid lines. In this plot, we
 assumed $\dot M_{\rm WR} = 2.2 \times 10^{-5} {\rm M}_{\odot} {\rm
 yr}^{-1}$, $\beta_{\rm WR}~=$ 1 and $\beta_{\rm O}~=$ 1 .}  
\label{fg:f3-6}
\end{figure}

  We calculate equation (\ref{eq1}) for three assumptions of the
  mass-loss rate ratio $\dot M_{\rm O}/\dot M_{\rm WR}$ $=$ 0.1, 0.04,
  and 0.02.  The $\beta$ model can reproduce the luminosity
  during observations A and B. However, it is truncated
  near periastron for all the assumptions since the wind momentum of the
  W-R star overwhelms that of the O star. This implies that the
  stagnation point almost arrives at the surface of the O star near
  observations C and D.  The observed (intrinsic, absorption-free) X-ray
  luminosity during observations C and D is then mostly produced near
  the surface of the O star.

One may think the radiative inhibition because of the stronger radiation
field of the O star (radiative breaking effect: \cite{stevens94}).
\citet{setia01} took the "eclipse spectra" ratio of the IUE spectrum
very close to periastron (phase 0.01) to a composite from phases 0.5 to
0.95 (figure 7 in \cite{setia01}).  The spectra shows red-shifted
absorption at 1300~${\rm km~s}^{-1}$ in the Si$\emissiontype{IV}$
1394--1403 doublet and 1400~${\rm km~s}^{-1}$ in the
C$\emissiontype{IV}$ 1548--1551 doublet. At this phase the O star behind
the W-R star was seen and the W-R wind was being looked through towards
the stagnation point. By correcting the velocity with the inclination of
54 degrees, the wind velocity was radiatively braked and was estimated
as about 2300~${\rm km~s}^{-1}$ before the collision with the O star
wind, i.e. about 80\% of its terminal velocity when it meets the O
stellar wind.

If the radiation breaking dominates the momentum balance in the
collision layer, the WC star winds lose its momentum and the cone angle
of the collision layer will be enlarged, making the X-ray luminosity
increased. The observed flux drop at the periastron indicates that the
radiative wind breaking is not a major effect to control the momentum
balance in the wind-wind collision layer (i.e., X-ray luminosity).

  In this condition, the mass-loss ratio $\dot M_{\rm O}/\dot M_{\rm
  WR}$ of 0.04 is good to explain the observed luminosities at
  observaion B.  If we adopted this mass-loss ratio and $\dot M_{\rm WR}
  \sim 2.2 \times 10^{-5} {\rm M}_{\odot} {\rm yr}^{-1}$, which
  reproduced the observed column densities at observations C and D (for
  the detail, refer to $\S$~\ref{subsec:mass-loss}), we obtain $\dot
  M_{\rm O} \sim 9 \times 10^{-7} {\rm M}_{\odot} {\rm yr}^{-1}$. This
  mass-loss rate for the O star is close to that recently reported by a
  number of analyses ($\dot M_{\rm O} = 8 \times 10^{-7} {\rm M}_{\odot}
  {\rm yr}^{-1}$; \cite{pittard06}, $\dot M_{\rm O} = 1 \times 10^{-6}
  {\rm M}_{\odot} {\rm yr}^{-1}$; \cite{fahed11}).

\subsection{Estimation of the W-R mass-loss rate}
\label{subsec:mass-loss}

We estimated the mass-loss rate of the W-R star from the observed column
density $N_{\rm He}$. The mass-loss rate for a spherically symmetric
wind is expressed as $\dot{M}_{\rm WR}$ $=$ 4$\pi$
$r^{2}~\rho(r)~v_{\rm{WR}}(r)$, in which {\it r} is the distance from
the center of the W-R star. Here, $\rho$ is the wind mass density.  In
addition the He-equivalent column density can be expressed as $N_{\rm
He} = \int \rho(r) dl \times \alpha / 4m_{\rm p}$ from the X-ray
emitting region to infinite distance along the line of sight ``$l$''
(cf. \cite{williams90}).  Here, $m_{\rm p}$ is the proton mass and
$\alpha$ is the ratio of helium mass to total mass in the W-R wind.  We
therefore apply for the mass-loss rate of the W-R star, whose column
density gives close agreement with the observed He-equivalent column
density.

The computed He-equivalent column density was obtained as follows. We
first calculated the stagnation point in consideration of the effect of
non-constant wind momenta, as shown in $\S$~\ref{subsec:flux}. Then, we
assumed that the X-ray is emitted from the stagnation point. Finally, we
calculated the absorption column density of the W-R stellar wind.  Here,
we used $\alpha = 0.4$, which was calculated by utilizing the elemental
abundances of table~\ref{tb:t3-2}. We adopted a simple beta law model
for the wind acceleration: $v(r)=v_{\infty}(1-R/r)^{\beta}$.  We used
the same values of $v_{\infty}$, $R$ and $\beta$ for the W-R and O stars
as those in $\S~\ref{subsec:flux}$. An orbital inclination angle
$i=120\degree$ was adopted \citep{dougherty05}.

Figure~\ref{fg:f3-6} shows a plot of the best-fit column density from
table~\ref{tb:t3-1}. We also plotted our calculated column density
through the W-R wind, in the case of $\dot M_{\rm WR} \sim 2.2 \times
10^{-5}~{\rm M}_{\odot} {\rm yr}^{-1}$. Here, we tested the three cases
with the mass-loss rate ratios of $\dot M_{\rm O}/\dot M_{\rm WR}$ $=$
0.1, 0.04, and 0.02. The calculated column densities for the $\beta_{\rm
WR}~=$ 1, 3 and 5 were much the same. In observation A, we set the
absorption column density caused by the W-R wind to zero. The absorption
column density by the wind increased as the binary approached
periastron. In the case of $\dot M_{\rm O}/\dot M_{\rm WR}$ $=$ 0.04,
and 0.02, the stagnation points at observations C and D were at the O
star surface.

We assumed that the wind-wind collision layer is concentrated in a small
region at the stagnation point. The good correspondence between the
model and the data, in observations C and D, indicates our assumption is
reasonable at least in the phases. On the other hand, at observation B,
there is a discrepancy by a factor of two between the data and the
model. The layer at observation B is largely extended away from the
stagnation point and the O star. During observation B, the X-rays
emitted from the wind-wind collision layer will suffer absorption at
different column density according to the different line of sight.

In the calculation of the wind absorption, we assumed values for the
stellar radius of the W-R star $R_{\rm{WR}}$, the inclination angle $i$
of the binary system and the C/He abundance ratio.  If the stellar
radius of the W-R star was in fact, for example, a half of our assumed
value, i.e., $R_{\rm{WR}}=~1~{\rm R}_{\odot}$, the absorption column at
periastron would be lower by roughly 10\% than the above-mentioned
value.  If the inclination angle was a five-degree larger than our
assumed value, the absorption column at periastron would be larger by
roughly 10\%.  If the assumed C/He abundance ratio was a half of it,
i.e., C/He~$=~0.2$, the absorption column at periastron would be larger
by roughly 20\%.  Or, if the ratio was doubled, i.e., C/He~$=~0.8$, the
column at periastron would be 20\% lower. As such these three parameters
are sensitive to $N_{\rm He}$. Accordingly the mass-loss rate derived
from $N_{\rm He}$ may have some considerable uncertainty.

\citet{williams90} derived the mass-loss rate $\dot M_{\rm WR} \sim 5.7
\times 10^{-5}~{\rm M}_{\odot}~{\rm yr}^{-1}$, which assumed a distance
of 1.3 kpc and made no allowance for clumping, from the radio flux at
the quiescent phase.  Allowing for the clumping having a filling factor
of 0.1 and the revised distance, the mass-loss rate $\dot M_{\rm WR}$
was determined as $ 2.6 \times 10^{-5}~{\rm M}_{\odot}~{\rm yr}^{-1}$.
By using X-ray absorption, we obtained $\dot M_{\rm WR} \sim 2.2 \times
10^{-5}~{\rm M}_{\odot}~{\rm yr}^{-1}$ at observations C and D near the
periastron.  Our value can coincide with that by using the radio flux.

\subsection{Upper limit of the hard band excess}
\label{subsec:hard-tail} 
We have detected a signal in the high-energy band above 10 {\rm keV}
(figure~\ref{fg:f3-3}). However, the Seyfert 2 Galaxy, IGR J20216+4359,
is located in the field of view of PIN/HXD.  The detected hard-tail
signal may be contaminated with the emission from IGR J20216+4359.

IGR J20216+4359 was discovered by \citet{bikmaev08} during their eleven
months observation. According to \citet{bikmaev08}, IGR J20216+4359 is
highly absorbed ($N_{\mathrm{H}} = 1.3 \times 10^{23}$
$\mathrm{cm}^{-2}$), when they fixed $\Gamma$ to 1.7, which is a typical
value of a Seyfert 2 Galaxy. By assuming the flux and the spectral shape
are the same as those in \citet{bikmaev08} through our observations, we
estimated the count rate of the contamination from IGR J20216+4359. The
model predicted count rate with HXD/PIN is dependent on the off-axis
angle of IGR J20216+4359. The count rates at observations A, B, C and D
are 0.013, 0.017, 0.016 and 0.016 counts per second in the 15--50~{\rm keV}
band, respectively.  The contaminations from IGR J20216+4359 accounts
for about 60 and 80\% of the excess count rates at observations A and
B, respectively. On the other hand, there are no excess count rates at
observations C and D. The count rates from off-axis IGR J20216+4359 at
observations C and D were 2.0 and 3.2 times larger than the observed
count rates in the 15--50~{\rm keV} band, respectively.  This suggests that
the flux of IGR J20216+4359 were varied. Since the flux of IGR
J20216+4359 was varied, we cannot constrain the origin of the hard band
excess in observations A and B. Therefore, we regard the flux detected
at observation B is an upper limit of the hard band excess from WR~140
($L_{\mathrm X} < 8\times10^{33}~\rm{erg}~\rm{s}^{-1}$, see
table~\ref{tb:t3-5}). {\it ASTRO-H} and {\it NuSTAR}, which have hard
X-ray imaging instruments with high sensitivity, will be useful to
verify the existence of the hard band excess.

In the colliding-wind region, first-order diffusive shock acceleration
results in the production of a power-law spectrum with index 2 for
electrons (e.g., \cite{bell78}, \cite{pollock87}, \cite{eichler93} and
\cite{white95}).  The best-fit photon index ($\Gamma \sim$ 1.7) indeed
approximates a photon index for non-thermal electron energy
distribution. \citet{pittard06} discuss the photon index of the
non-thermal emission from WR~140. They expected that the photon index is
1.7 above the energy of 1 MeV. Our fitting results are consistent with
it, although our results have a large error. According to
\citet{white95}, the luminosity ratio of inverse Compton to synchrotron
radiation can be written as ${L_{\rm{syn}}}/{L_{\rm{IC}}} = 840 {\it
B}^{2}{r_{\rm O}^{2}}/{L_{\rm{O}}}$, where the magnetic field ${\it B}$
is expressed in \rm{G}, $r_{\rm O}$ is the distance from the
colliding-wind zone to the O star in {\rm AU}, and $L_{\rm{O}}$ is the O
star luminosity in $L_{\odot}$ units.  For example, during observation
B, if we suppose $L_{\rm{IC}}<1\times10^{34}~\rm{erg}~\rm{s}^{-1}$ (see
table~\ref{tb:t3-5}),
$L_{\rm{syn}}=4.6\times10^{29}~\rm{erg}~\rm{s}^{-1}$ for a distance of
1.67 kpc (phase$=$0.93; \cite{dougherty05}),
$L_{\rm{O}}=5.7\times10^{39}~\rm{erg}~\rm{s}^{-1}$ \citep{dougherty05},
and $r_{\rm O} \sim 0.3~\rm{AU}$, the total value of the magnetic field
for observation B is $\it{B}>$1~{\rm G}. Because of the difficulty of
measuring the magnetic field of the W-R stellar surface, the value of
the magnetic field around the W-R star will be helpful to study the
evolution of the massive star.

 \section{Summary}
 We have observed the colliding-wind binary WR~140 around the 2009 periastron
 passage with the {\it Suzaku} satellite. The following are the salient results.

 \begin{itemize}
  \item [A.] We discovered a cool component. The absorption to this
	component is smaller than that to the other components. We infer
	that this component extends out farther than the other
	components does. 
	Arguably this component may represent a transitional phase from
	the compressed hot gas to dust formation.
	
  \item [B.] The luminosity of the hot component is not inversely
	proportional to the binary separation. This discrepancy may be
	explained if the O-star wind collides with the W-R star wind
	before it has reached its terminal velocity, leading to a
	reduction in its wind momentum flux. This interpretation needs
	to be verified by future theoretical modelings.
	
  \item [C.] As WR~140 approaches periastron, the column density of the
	hot component increases. The column densities at near periastron
	are reproduced well with a simple model, in which the absorption
	is occurred in the W-R wind and the X-ray emitting region is
	fairly compact.

  \item [D.] We detected a hard X-ray signal above 10 {\rm keV} in the
	HXD/PIN data. However, we could not eliminate the possibility of
	contamination from the background source IGR J20216+4359, 17.3
	arcmin north-east of WR~140. The PIN data gives the upper-limit
	for the X-ray luminosity of WR~140,
	8$\times10^{33}~\rm{erg}~\rm{s}^{-1}$.

 \end{itemize}

 We thank the referee for their help in improving the quality of this
 paper. We also thank M. Sakano for his invaluable help. This research
 has made use of data and/or software provided by the High Energy
 Astrophysics Science Archive Research Center (HEASARC), which is a
 service of the Astrophysics Science Division at NASA/GSFC and the High
 Energy Astrophysics Division of the Smithsonian Astrophysical
 Observatory.  Y.\,S. acknowledges financial support from the Japan
 Society for the Promotion of Science.  Y.\,T. and Y.\,M. acknowledge
 support from the Grants-in-Aid for Scientific Research (numbers
 20540237, 21018009 and 23540280) by the Ministry of Education, Culture,
 Sports, Science and Technology.  K.\,H. is grateful for financial
 support by the NASA's Astrobiology Institute (RTOP 344-53-51) to the
 Goddard Center for Astrobiology.  A.\,F.\,J.\,M. is grateful for
 financial aid to NSERC (Canada) and FQRNT (Qu\'ebec).  P.\,M.\,W. is
 grateful to the Institute for Astronomy for continued hospitality and
 access to the facilities of the Royal Observatory.  J.\,M.\,P. would
 like to thank the Royal Society for funding a University Research
 Fellowship.  This research has made use of NASA's Astrophysics Data
 System.

\end{document}